# Symmetry-designed $BiFeO_3$ single domain spin cycloid for efficient spintronics

Pratap Pal,[1,†] Jonathon L. Schad,[1,†] Anuradha M. Vibhakar,[2] Shashank Kumar Ojha,[3] Sajid Hussain,[4] Gi-Yeop Kim,[5] Saurav Shenoy,[6] Fei Xue,[6] Kaushik Das[7], Yogesh Kumar[8], Paul Lenharth,[1] A. Bombardi[2], Sayeef Salahuddin[7], Roger D. Johnson,[2,9] Si-Young Choi,[5] Mark S. Rzchowski,[10] Long-Qing Chen,[6] Ramamoorthy Ramesh,[3,4,11,12] Paolo G. Radaelli,[13] and Chang-Beom Eom[1*]

[1]Department of Materials Science and Engineering, University of Wisconsin-Madison, Wisconsin 53706, USA.
[2]Diamond Light Source Ltd, Harwell Science and Innovation Campus, Didcot, OX11 0DE, United Kingdom.
[3]Rice Advanced Materials Institute, Rice University, Houston, TX, 77005, USA.
[4]Department of Materials Science and Engineering, University of California, Berkeley, CA, 94720, USA.
[5]Department of Materials Science and Engineering, Pohang University of Science and Technology, Pohang, Gyeongbuk 37673, Republic of Korea.
[6]Department of Materials Science and Engineering, The Pennsylvania State University, University Park, Pennsylvania 16802, USA.
[7]Department of Electrical Engineering and Computer Sciences, University of California, Berkeley, CA 94720, USA.
[8]Materials Sciences Division, Lawrence Berkeley National Laboratory, Berkeley, CA, 94720, USA.
[9]Department of Physics and Astronomy, University College London, London, WC1E 6BT, United Kingdom.
[10]Department of Physics, University of Wisconsin-Madison, Wisconsin 53706, USA.
[11]Department of Materials Science and Nano Engineering, Rice University, Houston, TX, 77005, USA
[12]Department of Physics, University of California, Berkeley, CA, 94720, USA
[13]Clarendon Laboratory, Department of Physics, University of Oxford, Oxford OX1 3PU, United Kingdom.

*Corresponding author: ceom@wisc.edu
†These authors contributed equally.

## Abstract

Deterministic control of coupled ferroelectric and antiferromagnetic orders remains a central challenge in multiferroics, limiting their integration into functional magnetoelectrics and magnonic-devices. $(111)_{pc}$ $BiFeO_3$ with a robust single spin cycloid, offers direct magnetoelectric-coupling and a platform for efficient spin transport, yet multi-magnetic domains and ferroelectric-fatigue have prevented reproducible control. Here, we show that anisotropic-compressive in-plane strain stabilizes a single antiferromagnetic domain with unique spin-cycloid vector, by breaking the symmetry of the $(111)_{pc}$ plane. Epitaxial $BiFeO_3$ films grown on orthorhombic $NdGaO_3$ $(011)_o$ $[(111)_{pc}]$ substrates impose the required anisotropic in-plane strain and stabilizes single antiferromagnetic domain, as confirmed through direct imaging with scanning NV microscopy and non-resonant-x-ray-magnetic-scattering. Remarkably, these engineered films exhibit deterministic and non-volatile 180° switching of ferroelectric and single antiferromagnetic domains over 1,000 cycles. The monodomain state also enables anisotropic and threefold enhanced magnon transport with reduced scattering. Thus, symmetry-designed $(111)_{pc}$ monodomain $BiFeO_3$ offers a robust platform for advanced magnetoelectric and magnonic applications.

## Introduction

Multiferroic materials with strong magnetoelectric coupling enable electric-field control of magnetism, offering an energy-efficient pathway for next-generation spintronic and magnonic devices (*1–6*). Among these, $BiFeO_3$ (BFO) is the most extensively studied single-phase room-temperature magnetoelectric multiferroic(*7–10*), with coupled antiferromagnetic and ferroelectric orders. However, its optimal use is constrained by complex domain structures(*11*) of ferroelectric polarization, ferroelastic distortions and antiferromagnetic spin cycloid (fig. S3). BFO exhibits a rhombohedral (R3c) crystal symmetry, resulting in a total of 24 domain variants (4 ferroelastic × 2 ferroelectric × 3 antiferromagnetic)(*12–14*) complicating the study of magnetoelectric-switching, causing leakage(*15*) and polarization fatigue(*16*). Multiple antiferromagnetic domains hinder efficient magnon transport(*17*, *18*) though necessary for spintronic applications. Therefore, achieving a single-domain BFO thin film is desirable for electric-field-controlled magnetoelectric device applications.

Single domain control has been attempted in $(001)_{pc}$ BFO ("$_{pc}$" stands for pseudocubic) through substrate miscut(*11*, *19*). Controlling antiferromagnetic domains remains challenging as they respond weakly to external perturbations(*20*). Additionally, the polarization switching is dominated by 71° ferroelastic domain switching that is prone to back-switching and loss of retention(*21*). In contrast, $(111)_{pc}$ BFO allows polarization reversal within a single ferroelastic domain(*16*). In principle, a 180° reversal of the weak local magnetization (which changes sign with the same periodicity as the cycloid of the antiferromagnetic domain) is predicted to be achievable in $(111)_{pc}$ BFO under electric field switching(*22*), whereas it is limited in the $(001)_{pc}$ configuration(*23*, *24*). Although these factors should position $(111)_{pc}$ BFO as an ideal platform for integration into magnetoelectric devices, the three degenerate antiferromagnetic domains (Figs. 1**A** and **B**) lead to significant disadvantages for deterministic magnetic switching(*25*, *26*). Moreover, films in this orientation have been plagued by electrical breakdown and polarization fatigue attributed to the non-deterministic multi-path ferroelectric switching process arising from multiple domains(*16*, *27*). Consequently, stabilizing single antiferromagnetic domain in $(111)_{pc}$ oriented thin-film is a key objective in BFO research.

In a previous work(*27*) we employed substrate-induced anisotropic strain as an effective means to realize the antiferromagnetic spin cycloid with a definite propagation *k*-vector by growing epitaxial $(111)_{pc}$ BFO on $TbScO_3$ $(011)_O$ substrate(*27*) (similar results were also obtained using $DyScO_3$ $(011)_O$(*28*)) ("$_O$" stands for orthorhombic). These substrates impose tensile and compressive strains along $[11\text{-}2]_{pc}$ and $[1\text{-}10]_{pc}$ respectively, stabilizing a spin cycloid with a single propagation vector along the compressive strain direction. However, reliable electric field switching could not be achieved by this strategy. Although 95% of a single ferroelastic domain was obtained, minority domains persisted (~5%) along the direction of tensile strain and expanded to ~50% after one electrical switching from down to up polarization state(*27*). Thus, such a strategy poses a serious limitation toward deterministically switchable magnetoelectric device applications. Hence, we present a new and creative approach using a different anisotropic strain that robustly stabilizes single-domain state in $(111)_{pc}$ BFO even under multiple polarization reversal.

We replace the tensile strain with a compressive one while still maintaining the strong anisotropy (necessary for single spin cycloid) with a $NdGaO_3$ $(011)_O$ substrate (-2.65% along $[1\text{-}10]_{pc}$ and -2.0% along $[11\text{-}2]_{pc}$]) as seen in Figs. 1**B** and 1**C** (fig. S4). We demonstrate that anisotropic-compressive strain (Fig. **1C**) can lift the degeneracy of equivalent monoclinic domains(*25*, *27*, *29*) giving rise to single antiferromagnetic domain. Using real-space NV microscopy imaging and non-resonant X-ray magnetic scattering techniques, we directly probe the stabilization of a single spin-cycloid, with deterministic switching by electric field over 1000 times. Notably, our films exhibit a reliable and non-volatile polarization switching over one hundred thousand cycles due to the 180° single step ferroelectric switching (Fig. 1**D**), which is unique for this platform. We understand this microscopically with theoretical phase field calculations of domain reversal. This anisotropic-compressive strain design demonstrates that monodomain $(111)_{pc}$ BFO circumvents major challenges of reliable and nonvolatile electric field writing and enables efficient magnon spin transport over a multi-antiferromagnetic domain BFO, which is promising for potential magnetoelectric spintronic applications.

## Results and Discussions

### Anisotropic-compressive strain designed single domain $(111)_{pc}$ $BiFeO_3$

To realize a single spin-cycloid in $(111)_{pc}$ BFO with robust switching, we employed phase field calculations to probe the effect of anisotropic-compressive strain (Materials and Methods). The polarization is coupled to the antiferromagnetic order $\vec{L}$- through the Lifshitz invariant, that drives the incommensurate spin cycloid(*30–32*). For isotropic strain ($SrTiO_3$ (111)), calculations reveal three degenerate cycloid variants, whereas anisotropic-strain ($NdGaO_3$ $(011)_O$) lifts this degeneracy, stabilizing single variant (Figs. 1**B** and 1**C**). This is like previously observed anisotropic-tensile strain using Tb(Dy)$ScO_3$ substrate(*27*, *28*) But, unlike the previously explored tensile strain, where polarization switching seeds large minority domains, the compressive strain platform maintains a robust single domain state during electrical switching (Fig. 1**D** and fig. S4). Thus, $NdGaO_3$ $(011)_O$ substrate provides a robust route to stabilize single cycloid domain in $(111)_{pc}$ BFO with deterministic switching, overcoming limitation of tensile strain-based approaches(*27*, *28*).

Guided by our hypothesis, we grew high-quality $(111)_{pc}$ BFO epitaxial thin films on orthorhombic $NdGaO_3$ $(011)_O$ and (111) $SrTiO_3$ substrates (figs. S5 and S6) with $SrRuO_3$ bottom electrodes using off-axis magnetron sputtering. $NdGaO_3$ $(011)_O$ is expected to impose anisotropic-compressive in-plane strain (~2.65% along $[1\text{-}10]_{pc}$ and ~2.0% along $[11\text{-}2]_{pc}$), unlike the isotropic strain from $SrTiO_3$ substrates(*33*) or the anisotropic-tensile strain from Tb(Dy)$ScO_3$(*27*, *28*). Reciprocal space mapping (RSM) reveals that BFO on $SrTiO_3$ retains macroscopically high symmetry, while films on $NdGaO_3$ $(011)_O$ exhibit reduced in-plane symmetry at all length scales (fig. S7), consistent with our thermodynamic prediction for stabilizing single antiferromagnetic domain. Scanning transmission electron microscopy (STEM) further reveals BFO on $NdGaO_3$ $(011)_O$ forms a single ferroelastic domain (fig. S8) with atomically sharp interfaces (Figs. 2**B** and 2**C**)(*34*) despite the large lattice mismatch. Quantitative strain mapping confirms larger compressive strain along the $[1\text{-}10]_{pc}$ ($\varepsilon_{AA}$~2.75 %) than along ($[11\text{-}2]_{pc}$ ($\varepsilon_{BB}$~2.2 %) (Figs. 2**D** and 2**E**) validating the anisotropic-compressive strain state. Thus, $NdGaO_3$ $(011)_O$ offers a robust desired platform to control antiferromagnetic domains and enable deterministic switching in $(111)_{pc}$ BFO.

### Single variant spin cycloid in monodomain (111) $BiFeO_3$ thin films

Reciprocal space mapping using synchrotron x-ray diffraction revealed that $(111)_{pc}$ BFO on $NdGaO_3$ $(011)_O$ adopts a pure single $r_1$ ferroelastic domain state (fig. S23C), with no minority domains, which is in sharp contrast to films on scandate substrates that invariably host multiple variants (e.g., $TbScO_3$(*27*)and fig. S24 for $DyScO_3$). Piezoresponse force microscopy (PFM) further confirmed a uniform downward-directed ferroelectric polarization stabilized by the $SrRuO_3$ bottom electrode (fig. S9).

Using scanning NV diamond microscopy (Fig. 3**A**), we directly visualized the antiferromagnetic order by mapping the weak surface stray magnetic field caused by Dzyaloshinskii–Moriya interaction-driven spin-density wave(*35*, *36*). Typical dual iso-*B* NV scans show a striking contrast between isotopically strained BFO on $SrTiO_3$ and anisotropically strained films on $NdGaO_3$ (Figs. 3**B** and 3**D**). BFO on $SrTiO_3$ displays multiple spin-cycloid propagation vectors (Fig. 3**B**), accompanied by complex domain walls and intriguing topological defects(*37*). First Fourier transform (FFT) analysis (Fig. 3**C**) further confirms such degeneracy. In contrast, the BFO film on $NdGaO_3$ $(011)_O$ reveals a well-ordered periodic pattern of single spin-cycloid (Figs. 3**D** and 3**E**) with a propagation vector along $[1\text{-}10]_{pc}$ (fig. S14), which is further validated by full-B NV measurements (fig. S15) that quantifies the modulation of the local stray magnetic field(*36*). Such a type I spin cycloid resides in the $(11\text{-}2)_{pc}$ plane, containing $[111]_{pc}$ and $[1\text{-}10]_{pc}$, and couples to polarization along [-1-1-1]$_{pc}$(*38*). Importantly, the presence of spin textures featuring topological defects(*37*, *39*) is wide and complex for BFO films in case of $SrTiO_3$ (fig. S16), while that is largely absent with only sparse line defects in case of anisotropic and compressively strained $NdGaO_3$ (fig. S17). Thus, these results demonstrate that anisotropic-compressive strain breaks spin-cycloid degeneracy, stabilizes an antiferromagnetic monodomain state, and provides a pathway toward deterministic magnetoelectric switching in multiferroics.

While the stabilization of single ferroelastic and ferroelectric domains is well understood, the stabilization of antiferromagnetic monodomains (single variant spin cycloid) is less straightforward(*40*). To address this, we conducted thermodynamic phase-field simulations implementing the anisotropic-compressive strain parameters imposed by $NdGaO_3$ $(011)_O$. Our results reveal that despite starting with initial random noise, only one type of antiferromagnetic domain with a distinct spin-cycloid propagation direction (Fig. 3**F**) gets stabilized, consistent with the experiment. The driving mechanism is strong magnetoelastic coupling(*41*) induced by anisotropic strain. Interestingly, intermediate simulation stages show spin textures with topological defects (fig. S18) like those observed experimentally in scanning NV images. These transient defects gradually diminish and lead to an ordered single spin-cycloidal pattern. Together, these results demonstrate that anisotropic-compressive strain provides a robust pathway for stabilizing an antiferromagnetic monodomain in $(111)_{pc}$ BFO.

**Deterministic and non-volatile switching of magnetoelectric-multiferroic orders**

To probe the stability of complete monodomain states under electric field switching, we combined PFM, scanning NV imaging, and non-resonant X-ray magnetic scattering (NXMS). An apparent hysteresis of the vertical piezoresponse is observed, indicating perfect 180° polarization switching from down (↓) to up (↑) (fig. S9). Similarly, a box-in-a-box pattern (fig. S10) was written with the PFM tip and scanning NV images were recorded (fig. S19), which show a strong electric field response, with switching from a down (↓) to up (↑) polarization state. Similar observations had been previously attributed to the emergence of a canted AFM-like state with uniform magnetization, at least near the *surface* of the film(*40*). Notably, after completing one full cycle of switching (down-up-down), a robust single variant of the spin-cycloid (fig. S19) is restored. Due to the slow switching of polarization by PFM tip, there can be surface charge built up(*42*), causing strong signals(*43*) interfering the detection of spin-cycloids via NV microscopy.

To directly probe this effect, we used an Au top electrode to switch the ferroelectric polarization homogeneously and subsequently removed it (Materials and Methods) for NV imaging. First, we confirm a robust 180º polarization reversal via PFM probe (Fig. 4**A,** fig. S20). Interestingly, scanning NV image on a switched region beneath the electrode revealed a persistent robust single spin-cycloid domain with the same propagation vector as the unswitched region (Fig. 4**B**, fig. S21), demonstrating robust and reproducible magnetoelectric coupling.

To study the effect of repetitive fast switching (much faster than PFM), we applied a square electric field pulse to reverse the ferroelectric polarization by 180° over a hundred thousand cycles. Whereas BFO on $SrTiO_3$ (111) rapidly degrades after just a few thousand cycles(*16*), BFO on $NdGaO_3$ $(011)_O$ remains robust with negligible loss (Fig. 4**C**). Such a response from BFO on anisotropic-compressive $NdGaO_3$ $(011)_O$ outperforms previous anisotropic-tensile $Tb(Dy)ScO_3$ $(011)_O$ substrates (Fig. 4**C**), as tensile strain promotes undesired minority domain formations (Fig. 1**D**)(*27*). Thus, the polarization-fatigue-free response from the $(111)_{pc}$ monodomain BFO system with record low leakage compared to its multidomain variants (figs. S12 and S13) represents a critical step toward deterministic information writing in devices. Notably, this deterministic switching is non-volatile, unlike (001) monodomain BFO(*21*), which remains stable for over thousands of hours (fig. S10).

The effect of electric field switching (fig. S22) on magnetic ordering in the *bulk* of our films was probed by NXMS measurements using synchrotron x-ray diffraction on satellites of the $(009)_h$ reflection ("*h*" denotes the hexagonal setting), corresponding to long-range incommensurate (cycloidal) magnetic ordering. Reciprocal-space mapping of a 200μm × 50 μm area of the as grown sample revealed a single pair of magnetic satellite peaks (Fig. 4**D**) at $(009)_h \pm \mathbf{k}$, aligned along the $[100]_O$ ($[1\text{-}10]_{pc}$) substrate direction, confirming a single antiferromagnetic cycloidal domain associated with the major $r_1$ ferroelastic domain and a coherent magnetic structure, consistent with the NV imaging. Mapping of a 75μm × 50μm area, electrically switched by 1,000 times and left in the up (↑) polarization state (fig. S22), yielded the same two magnetic satellite peaks (Fig. 4**E**) with similar peak-widths (Figs. 4**F** and 4**G**). This result demonstrates the persistence of a single antiferromagnetic cycloidal domain in both as-grown ($\downarrow \vec{P}$) and switched ($\uparrow \vec{P}$) polarization states, excellently corroborating the NV microscopy results. In a previous work(*27*), we employed polarized NXMS to demonstrate the *polarity (chirality)* of the monodomain cycloidal state in $(111)_{pc}$ BFO grown on $TbScO_3$ $(011)_O$ *switches,* upon electrical switching (our previous $TbScO_3$ data are reproduced in fig. S25). Although not performed here, identical switching is expected in case of films on $NdGaO_3$, due to strong magnetoelectric coupling in BFO (fig. S27). The fraction of the ferroelastic domain associated with the single spin-cycloid remains, on average, greater than 99.0% after multiple switching events (fig. S23), establishing a single magneto-structural variant that persists under extensive electric field cycling. These results highlight how interfacial strain (although relaxed in overall 1 μm thick film) manipulation stabilizes robust ferroelastic, ferroelectric, and antiferromagnetic monodomains, representing a major advance in achieving electrically switchable, fatigue-free (111) BFO thin films.

**Mechanism of deterministic switching in $(111)_{pc}$ $BiFeO_3$**

We next examine why BFO grown on $NdGaO_3$ $(011)_O$ is particularly suited for robust single domain multiferroic order. In $(111)_{pc}$ BFO, 180° polarization switching under an out-of-plane applied electric field can involve intermediate steps, such as 71° followed by 109° switching (fig. S26). In case of BFO on $SrTiO_3$, initial 71° switching can occur via multiple pathways, producing head-to-head (or tail-to-tail) charge domain walls that drive polarization fatigue(*44*, *45*). By contrast, large anisotropic-compressive strain imposed by orthorhombic $NdGaO_3$ $(011)_O$ breaks the symmetry, favoring a single-step 180° switching path, as confirmed by phase field simulations (Fig. 1**D** and fig. S26). This suppresses formation of charged domain walls providing a fatigue-free ferroelectric switching response over $>10^5$ cycles, in contrast to $DyScO_3$ or $TbScO_3$ with anisotropic tensile-compressive strain, which promotes minority domain formation and degradation (fig. S26). The large anisotropic-compressive strain in the present case, thus stabilizes a robust, single-domain $(111)_{pc}$ BFO thin films, marking a significant advancement in the field.

**Efficient magnon spin transport**

Armed with the robust single-variant spin-cycloid with deterministic switching, we performed a nonlocal magnon spin transport on $(111)_{pc}$ BFO films grown on differently strained substrates. A device structure with Pt source and detector pads were fabricated aligning along parallel and perpendicular to the spin cycloid propagation direction ($[1\text{-}10]_{pc}$) (Figs. 5**A** and **B)**. An electrical current sent through the source wire induces a temperature gradient in BFO, generating thermally excited spin waves via the spin-Seebeck effect. These magnons were detected via the inverse spin-Hall effect (ISHE) induced DC voltage ($\Delta V_{NL,2\omega}$) at the detector 200 nm from the source (fig. S28). While (111) BFO grown on $SrTiO_3$ shows isotropic magnon transport due to disordered and degenerate spin-cycloids, the same on $NdGaO_3$ $(011)_O$ exhibits highly anisotropic transport with higher efficiency when $\Delta V_{NL,2\omega}$ measured along the direction of spin-cycloid propagation. As the ISHE voltage $\Delta V_{NL,2\omega}$ serves as a direct indicator of the efficiency of spin-wave conduction, these results indicate that the presence of a single spin-cycloid reduces scattering and enhanced spin-wave conduction. These results highlight the unique advantage of anisotropic-compressive strain in stabilizing robust single-domain spin-cycloids for practical spintronic devices.

**Conclusions and Outlook**

We have demonstrated the stabilization of robust monodomain states (single ferroelastic, ferroelectric, and antiferromagnetic) and deterministic single-step switching in $(111)_{pc}$ $BiFeO_3$ thin films through anisotropic-compressive strain and symmetry engineering (Table I). Notably, we achieved nearly fatigue-free, deterministic electric field switching sustained over one hundred thousand cycles, employing such an anisotropic-compressive strain strategy, which outperforms previously used either isotropic or anisotropic-tensile strain approaches. This electric field writing is non-volatile, retaining information for thousands of hours, an essential feature for electric field-based memory applications.

Since the ferroelectric polarization ($P$) is linearly coupled to magnetic polarity ($\lambda = r_{ij} \times (S_i \times S_j)$, where $S_i$ and $S_j$ are nearest-neighbor spins, and $r_{ij}$ is a vector connecting spins $i$ and $j$), switching $P$ with an applied electric field is expected to induce a corresponding reversal of the cycloidal polarity (fig. S27)(*26*, *46*). This opens the possibility of achieving 180° electric field switching of the magnetic moment in a thin ferromagnetic overlayer (e.g., Co(*24*)) on top of $(111)_{pc}$ monodomain $BiFeO_3$ through one-to-one exchange coupling (between single-cycloid antiferromagnetism and top ferromagnetism), an exciting direction for future research.

Additionally, recent advances in non-local spin transport via low-energy magnon modes show promise for the development of magnon based spintronic devices(*6*, *17*, *47*) In this context, $(111)_{pc}$ monodomain $BiFeO_3$ thin films offer a compelling platform for unconventional, anisotropic, and highly efficient spin transport in non-local devices. We have successfully demonstrated efficient magnon transport on our single domain $(111)_{pc}$ platform, which provides proof of concept of uniform single spin cycloid efficient magnon propagation in the case of $NdGaO_3$ $(011)_O$ substrate, that outperforms that for the homogeneous multi-spin cycloid phase on $SrTiO_3$ substrate. Thus, our design principle represents a significant step toward optimizing single ferroelectric and antiferromagnetic domains in $BiFeO_3$, advancing the development of low-power, electric field-switchable memory devices.

**Acknowledgments**

CBE acknowledges support for this research through a Vannevar Bush Faculty Fellowship (ONR N00014-20-1-2844), and the Gordon and Betty Moore Foundation's EPiQS Initiative, Grant GBMF9065. Ferroelectric measurement at the University of Wisconsin–Madison was supported by the US Department of Energy (DOE), Office of Science, Office of Basic Energy Sciences (BES), under award number DE-FG02-06ER46327. G.Y.K. and S.Y.C. acknowledge support from Korea Basic Science Institute (National Research Facilities and Equipment Center) Grant (2020R1A6C101A202) funded by the Ministry of Education. S.K.O is supported by the NSF-FUSE program. S.H. and R.R. acknowledge research sponsored by the Army Research Laboratory as part of the Collaborative for Hierarchical Agile and Responsive Materials (CHARM) under Cooperative Agreement Number W911NF-24-2-0100. The views and conclusions contained in this document are those of the authors and should not be interpreted as representing the official policies, either expressed or implied, of the Army Research Laboratory or the U.S. Government. The U.S. Government is authorized to reproduce and distribute reprints for Government purposes, notwithstanding any copyright notation herein. S.S. and L.Q.C. are supported by the US Department of Energy, Office of Science, Basic Energy Sciences, under Award Number DE-SC0020145 as part of the Computational Materials Sciences Program. We gratefully acknowledge Dr. Andrea Morales and QZabre, Switzerland, for providing the NV-based quantum scanning data presented in Fig. 4**B**.

**Author contributions**

P.P., J.S., and C.B.E. conceived the project. P.P. and J.S. carried out film growth, characterizations, and device fabrication of $BiFeO_3$/Pt samples. A.V., R.J., and P.G.R. performed the NXMS and synchrotron RSM measurements. S.K.O. and R. R. performed NV diamond microscopy and PFM imaging. S. S., F. X., and L. Q. C. performed the phase field calculations. G. K and S. J. C. carried out STEM measurements. S.H.

performed the non-local spin transport measurements. K.D. and Y.K. prepared the non-local devices using e-beam lithography. P.P. and C.B.E. wrote the manuscript with contributions from all other co-authors. C.B.E. led the whole project.

**Competing interests**

The authors declare no competing interests.

**Data availability**

The data that support the findings of this study are available from the corresponding author on reasonable request.

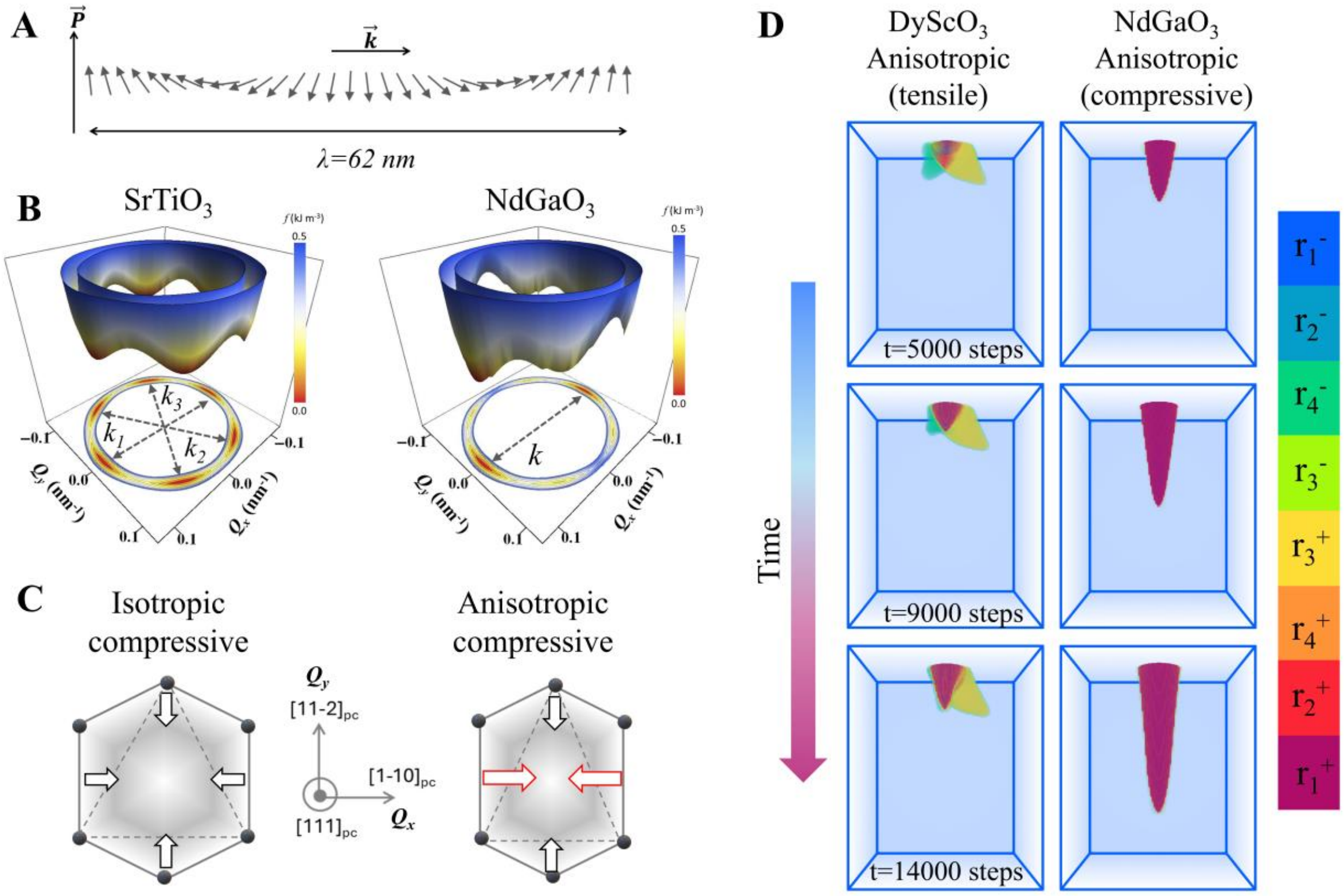

**Fig. 1. Anisotropic-compressive strain-designed AFM monodomain for deterministic switching. (A)** Schematic illustration of the spin cycloid in $BiFeO_3$, showing its alignment within the plane of polarization and propagation vector *k*. (**B)** Thermodynamic simulations showing three degenerate AFM domains under isotropic-compressive strain of 0.95% (e.g., $SrTiO_3$) and stabilization of a single type I spin-cycloid variant under anisotropic compressive strain, 2.65% along $[1\text{-}10]_{pc}$ and 2% along $[11\text{-}2]_{pc}$ directions (e.g., $NdGaO_3$). (**C)** Schematic illustration of the symmetry breaking of the $(111)_{pc}$ plane under different strains. **(D)** Phase field simulations of the deterministic 180° switching of $(111)_{pc}$ $BiFeO_3$ thin-films in the case of $NdGaO_3$ $(011)_O$ with *anisotropic-compressive* strain and in the case of *R*$ScO_3$ (*R*=Dy or Tb) $(011)_O$ with *anisotropic-tensile* strain (fig. S4).

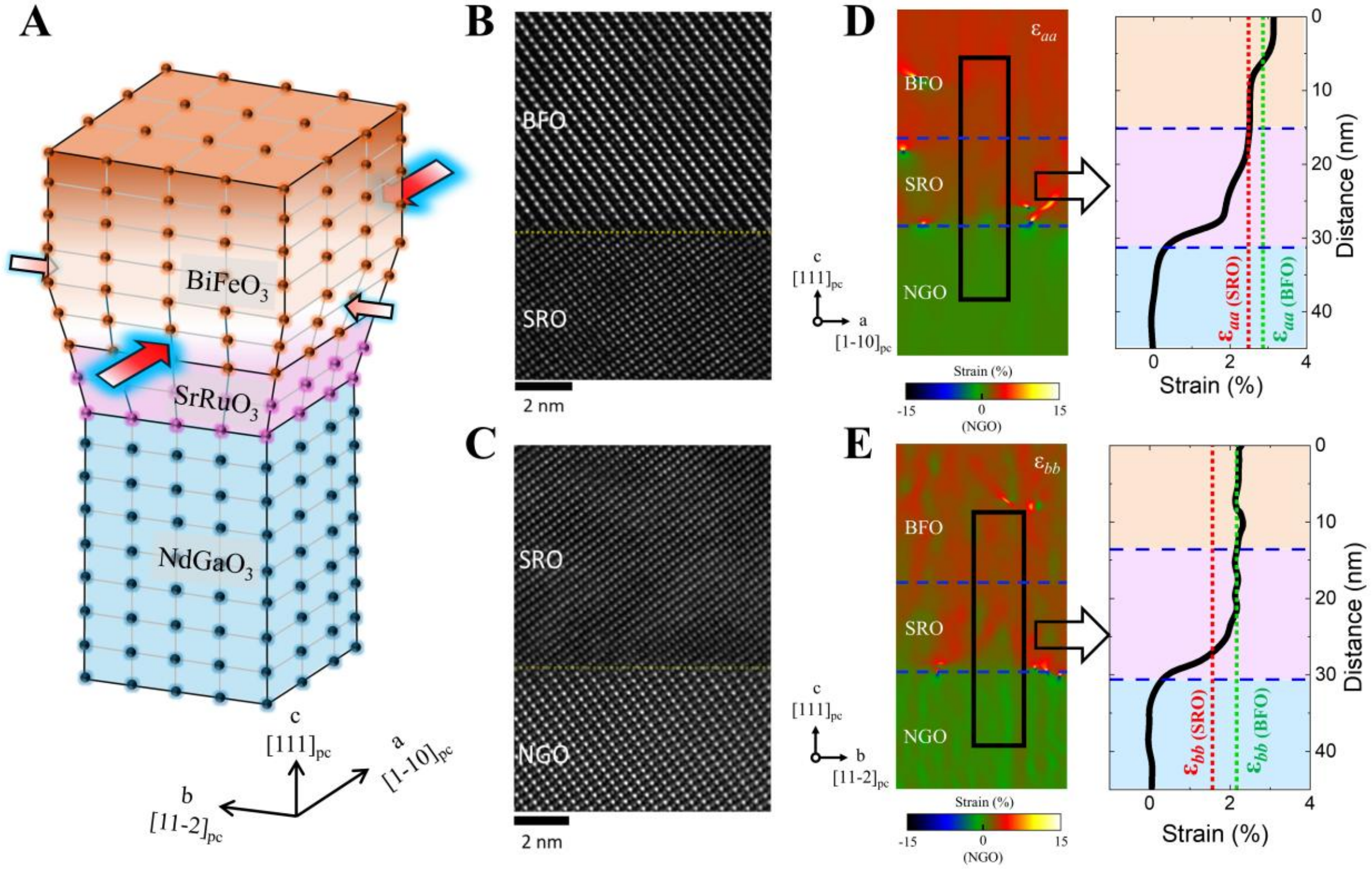

**Fig. 2. High-quality Epitaxial $(111)_{pc}$ $BiFeO_3$ with anisotropic in-plane strain.** **(A)** Schematic representation of the $BiFeO_3$ film showing anisotropic strain between *A* ($[1\text{-}10]_{pc}$) and *B* ($[11\text{-}2]_{pc}$) directions near the interface. (**B-C)** Cross-sectional high-resolution HAADF-STEM images along the $[1\text{-}10]_{pc}$ zone axis, revealing atomically sharp interfaces. (**D-E)** In-plane lattice strain maps ($\varepsilon_{AA}$ and $\varepsilon_{BB}$) and line profiles at the $BiFeO_3/SrRuO_3/NdGaO_3$ interface along the *A* and *B* projections, respectively. The in-plane lattice strain maps were calculated from HAADF-STEM images in fig. S8. In the line profiles, the red and green lines represent the lattice strain values of $BiFeO_3$ layers for each projection direction. The intensity profiles indicate that the initial $BiFeO_3$ layer is anisotropically strained, with greater strain along the *A* ($[1\text{-}10]_{pc}$) direction.

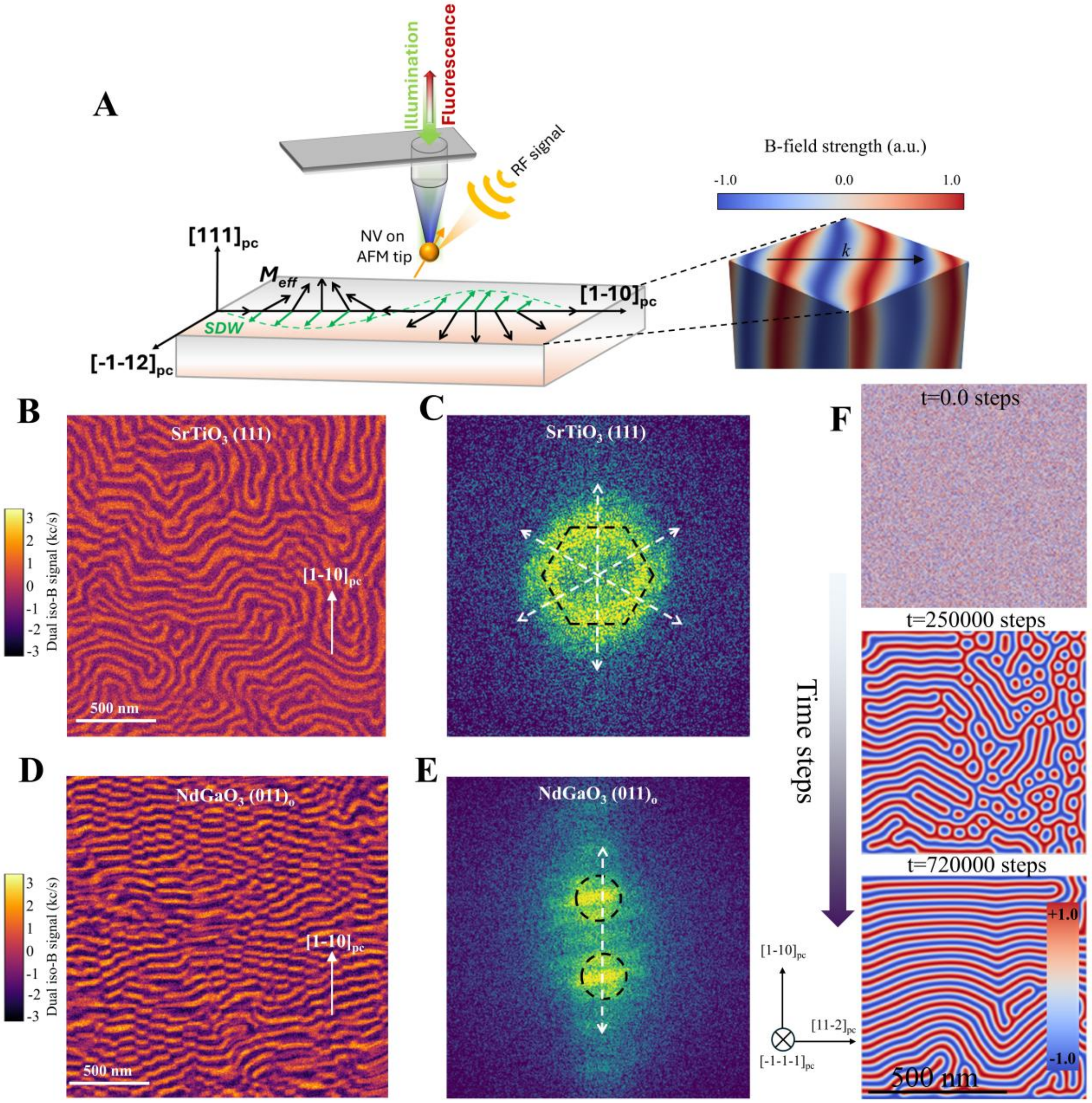

**Fig. 3. Real space imaging of antiferromagnetic domains. (A)** Schematic illustration of scanning NV diamond microscope setup used to probe the periodic modulation of stray magnetic field on the surface of $(111)_{pc}$ $BiFeO_3$. (**B)** Corresponding NV scan of the as-grown $BiFeO_3$ films on $(111)_{pc}$ $SrTiO_3$, showing a complex pattern with intriguing spin texture. (**C)** Its Fast-Fourier-Transform (FFT) image shows the apparent hexagonal pattern for three degenerate AFM domains. (**D-E**) Corresponding scanning images for $BiFeO_3$ on $NdGaO_3$ $(011)_O$ substrate, displaying much ordered pattern with single spin-cycloid. (**F)** Thermodynamic phase field simulation of the evolution of the AFM order and domain pattern for an initial configuration with random noise, 250,000 time steps and 720,000 steps. The presence of spin-texture from simulation indicates its origin due to thermal disorder during high-temperature thin-film growth.

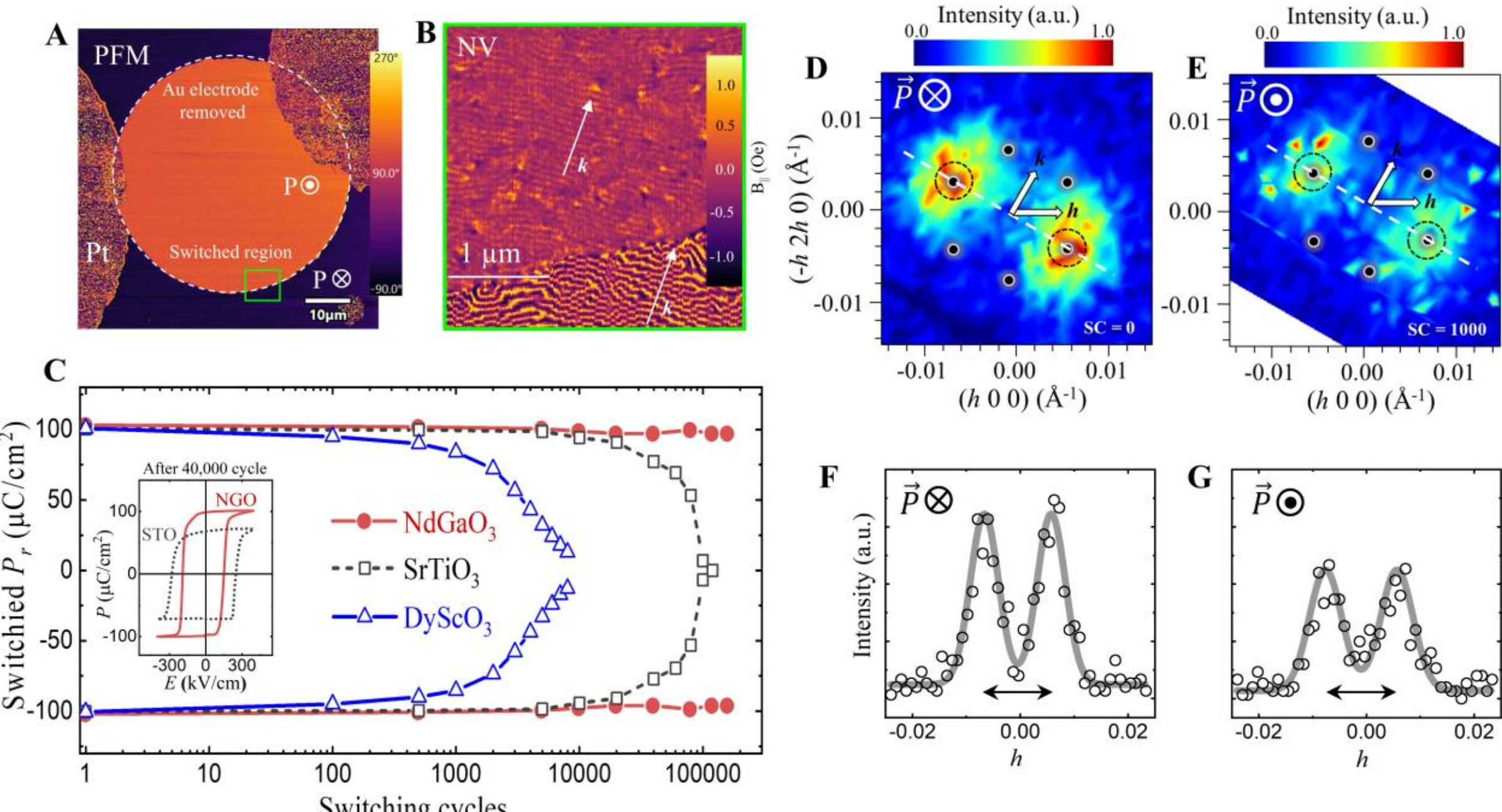

**Fig. 4. Deterministic electric-field switching of magnetoelectric order.** (**A)** PFM confirms the perfect 180º polarization switching from down (↓) to up (↑) state under an Au electrode that is removed after switching. (**B)** Corresponding NV microscopy image clearly showing the presence of a robust spin cycloid after polarization switching (fig. S21). (**C**) Ferroelectric fatigue measurement data for the BFO film on $NdGaO_3$ $(011)_O$, obtained using a square electric field pulse with an amplitude of 300 kV/cm and a frequency of 10 kHz. The inset to (**C**) showing the ferroelectric polarization data after 40,000 switching cycles. **(D-E)** NXMS scans near the $(009)_h$ reflection (specific to AFM domain) on the as-grown (down polarization state) and after 1,000 cycles of ferroelectric polarization switching (left in up polarization state) of BFO film on $NdGaO_3$ $(011)_O$. **(F-G)** Corresponding line scans across the magnetic peaks clearly exhibiting the presence of an AFM domain with single spin-cycloid propagation for both the as-grown (down $\vec{P}$ ) and after 1,000 cycles of ferroelectric polarization switching (left in up $\vec{P}$) states, highlighting the robustness single spin cycloid switching.

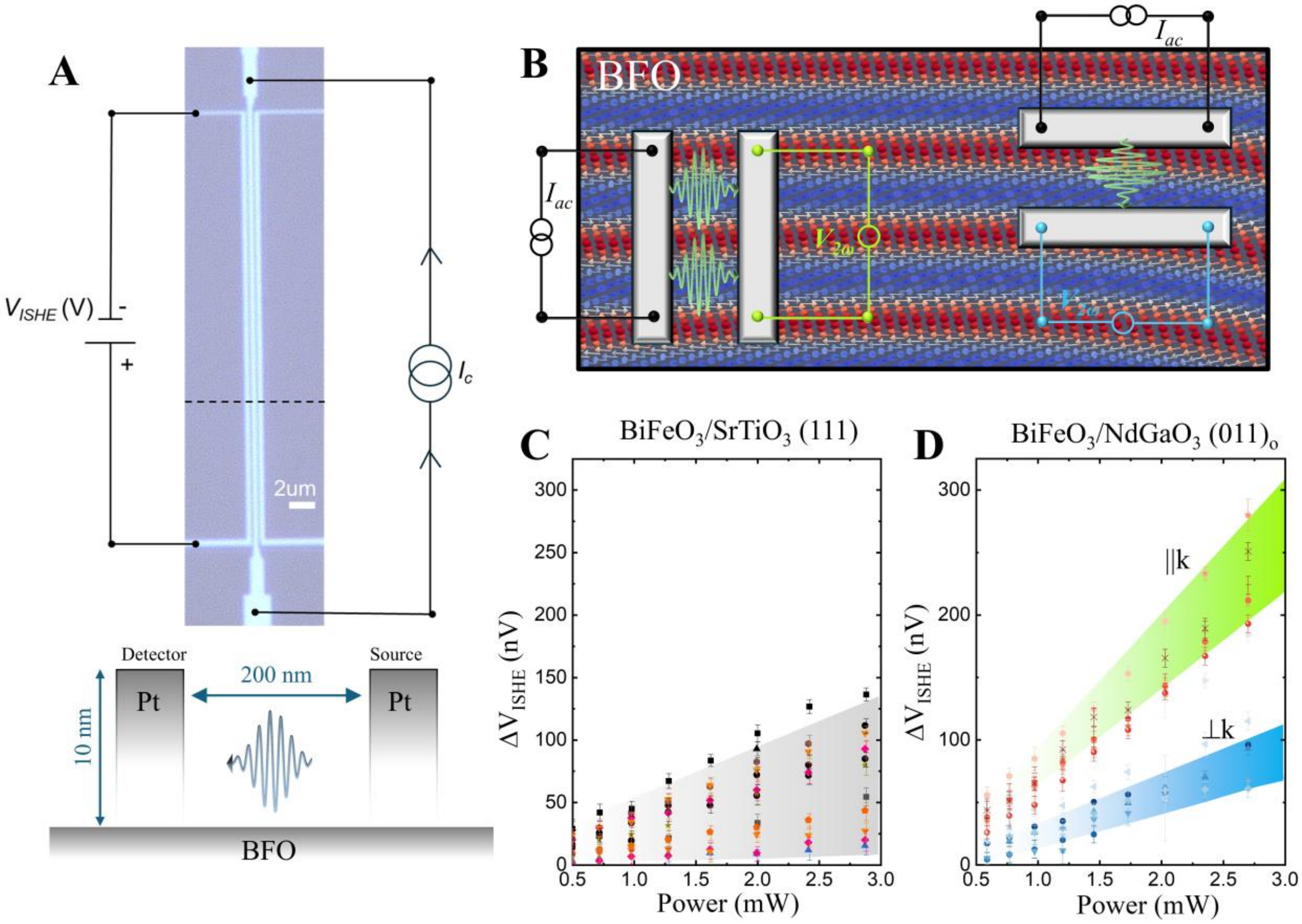

**Fig. 5. Non-local magnon transport device demonstration.** (**A**) Typical optical microscope image of the magnon device with a source and detector (made of 10 nm thick Pt) separation of 200 nm. (**B)** Schematic of the non-local spin transport device made on $(111)_{pc}$ $BiFeO_3$ with illustration of magnon transport along parallel (‖k) and perpendicular (⊥k) to the sing-spin cycloid propagation. (**C**) Nearly isotropic variation of the inverse spin-Hall effect (ISHE) voltage ($\Delta V_{ISHE}$) in case BFO on (111) $SrTiO_3$ with three degenerate antiferromagnetic domains. (**D**) Strong anisotropic variation of $\Delta V_{ISHE}$ with higher magnitude in case BFO on $(011)_O$ $NdGaO_3$ with single spin-cycloid propagation. $\Delta V_{ISHE}$ detected along the spin cycloid propagations is three times higher than that perpendicular to spin cycloid direction.

**Table 1. Comparison of the domain stability of $(111)_{pc}$ $BiFeO_3$ under electric field switching when grown on different substrates.**

| Substrates | Before switching ($P$↓) | | | After multiple switching ($P$↑) | | |
|---|---|---|---|---|---|---|
| | Ferroelastic | FE | AFM | Ferroelastic | FE | AFM |
| $SrTiO_3$ (111) (Isotropic compressive)(*16*) | Single | Single | Multiple | Multiple | Multiple | Multiple |
| $TbScO_3$ $(011)_O$ (Anisotropic tensile)(*27*) | Single | Single | Single | Multiple | Multiple | Multiple |
| $NdGaO_3$ $(011)_O$ (Anisotropic compressive) (This work) | Single | Single | Single | Single | Single | Single |

Supplementary Materials for

# Symmetry-designed $BiFeO_3$ single domain spin cycloid for efficient spintronics

Pratap Pal,[1,†] Jonathon L. Schad,[1,†] Anuradha M. Vibhakar,[2] Shashank Kumar Ojha,[3] Sajid Hussain,[4] Gi-Yeop Kim,[5] Saurav Shenoy,[6] Fei Xue,[6] Kaushik Das[7], Yogesh Kumar[8], Paul Lenharth,[1] A. Bombardi[2], Sayeef Salahuddin[7], Roger D. Johnson,[2,9] Si-Young Choi,[4] Mark S. Rzchowski,[10] Long-Qing Chen,[5] Ramamoorthy Ramesh,[3,4,11,12] Paolo G. Radaelli,[13] and Chang-Beom Eom[1*]

[1]Department of Materials Science and Engineering, University of Wisconsin-Madison, Wisconsin 53706, USA.
[2]Diamond Light Source Ltd, Harwell Science and Innovation Campus, Didcot, OX11 0DE, United Kingdom.
[3]Rice Advanced Materials Institute, Rice University, Houston, TX, 77005, USA.
[4]Department of Materials Science and Engineering, University of California, Berkeley, CA, 94720, USA.
[5]Department of Materials Science and Engineering, Pohang University of Science and Technology, Pohang, Gyeongbuk 37673, Republic of Korea.
[6]Department of Materials Science and Engineering, The Pennsylvania State University, University Park, Pennsylvania 16802, USA.
[7]Department of Electrical Engineering and Computer Sciences, University of California, Berkeley, CA 94720, USA.
[8]Materials Sciences Division, Lawrence Berkeley National Laboratory, Berkeley, CA, 94720, USA.
[9]Department of Physics and Astronomy, University College London, London, WC1E 6BT, United Kingdom.
[10]Department of Physics, University of Wisconsin-Madison, Wisconsin 53706, USA.
[11]Department of Materials Science and Nano Engineering, Rice University, Houston, TX, 77005, USA
[12]Department of Physics, University of California, Berkeley, CA, 94720, USA
[13]Clarendon Laboratory, Department of Physics, University of Oxford, Oxford OX1 3PU, United Kingdom.

*Corresponding author: ceom@wisc.edu
†These authors contributed equally.

**Supplementary Text 1**

**Materials and Methods**

1. Thin films sample growth, and characterization.

$(111)_{pc}$ monodomain $BiFeO_3$ thin films were grown on cubic $SrTiO_3$ (111) and orthorhombic $NdGaO_3$ $(011)_O$ substrates using RF magnetron sputtering. First, a 25 nm thick epitaxial $SrRuO_3$ (SRO) bottom electrode layer was deposited via 90° off-axis sputtering(*1*) at 600°C, followed by the growth of 1000 nm $BiFeO_3$ films using double-gun off-axis sputtering(*2*). The $SrRuO_3$ layer serves three purposes: it acts as a bottom electrode for out-of-plane device measurements, provides a depolarization field that orients the $BiFeO_3$ polarization downward, and serves as a buffer layer between the large lattice-mismatched $BiFeO_3$ and the substrate(*3*). The $BiFeO_3$ target used contains 5% excess $Bi_2O_3$ to compensate for bismuth volatility during thin-film deposition

2. Device fabrication

For ferroelectric switching, a ~10 nm Pt layer was patterned into 500 μm × 200 μm devices to serve as the top electrode, while the $SrRuO_3$ layer served as the bottom electrode.

For non-local magnon transport, 40 μm long and 400 nm wide metal stripes were patterned through Pt (10 nm) lift-off employing a 130 keV Crestec tool with a 100 keV electron beam, 600 micrometer field size, 60000 dots, and 200 pA dose current. The exposure base time is 1.2 microseconds, and the proximity effect correction factors are calculated using BEAMER software. Before patterning, we carried out spin-coating using PMMA 950 A2, at 3500 rpm for 30 seconds, followed by baking at 180ºC, for 60 seconds. After patterning, it was developed in 3:1 IPA: DI water for 60 seconds. This process was conducted at the Marvell Nanofabrication Laboratory at UC Berkeley.

3. Ferroelectric P vs E loop measurements

Ferroelectric polarization versus applied electric field measurements were carried out using a Precision Multiferroic Ferroelectric Tester from Radiant Technologies Inc. A vertical capacitor structure of $BiFeO_3$ was fabricated with ~10 nm Pt as the top electrode and ~30 nm $SrRuO_3$ as the bottom electrode. A triangular electric pulse was applied to measure the P vs E loop, while a square electric pulse with a frequency of 10 kHz and an amplitude of 300 kV/cm was used to switch the ferroelectric polarization over many thousands of cycles.

4. NXMS measurements

Synchrotron X-ray diffraction measurements, including structural characterization and non-resonant X-ray magnetic scattering (NXMS), were performed on the I16 beamline at Diamond Light Source (United Kingdom)(*4*), using a six-circle kappa diffractometer in reflection geometry. A 4.9 keV beam, off-resonance to all chemical elements in the sample, with a ~50 μm beam profile, was used for the NXMS measurements. A Pilatus area detector, APD, and room-temperature sample stage were used. The X-ray beam size at the sample was adjusted to approximately 50 μm × 50 μm for scanning.

5. Theoretical calculations

The influence of epitaxial strain on the antiferromagnetic order in the $BiFeO_3$ film was investigated via a thermodynamic analysis. Using order parameters describing the spontaneous polarization **P** and the antiferromagnetic vector **L,** an expression for the total free energy of a single ferroelectric domain was constructed. The total free energy includes four energy terms i.e., an antiferromagnetic anisotropic term, the antiferromagnetic exchange energy, the Lifshitz-type magnetoelectric coupling term and the elastic energy (*5*, *6*). The total free energy is given by

$$F = \int dV \left\{ K_1(L_1^2L_2^2 + L_2^2L_3^2 + L_3^2L_1^2) + K_2L_1^2L_2^2L_3^2 + A\sum_{i=1}^{3}(\nabla L_i)^2 + \gamma\mathbf{P}\cdot[\mathbf{L}(\boldsymbol{\nabla}\cdot\mathbf{L}) - (\mathbf{L}\cdot\boldsymbol{\nabla})\mathrm{L}] + \frac{1}{2}c_{ijkl}(\varepsilon_{ij} - \varepsilon_{ij}^0)(\varepsilon_{kl} - \varepsilon_{kl}^0) \right\}$$

Where $K_1$ and $K_2$ are the antiferromagnetic anisotropy constants, $A$ is the antiferromagnetic exchange constants, $\gamma$ is the coefficient for the magnetoelectric coupling term, $c_{ijkl}$ is the elastic stiffness tensor, $\varepsilon$ is the epitaxial strain of the film, and $\varepsilon^0$ is the stress-free strain. The stress-free strain $\varepsilon^0$ includes the contributions from electrostriction and magnetostriction, i.e., $\varepsilon_{ij}^0 = Q_{ijkl}P_kP_l + \lambda_{ijkl}L_kL_l$ $\varepsilon_{ij}^0 = Q_{ijkl}P_kP_l + \lambda_{ijkl}L_kL_l$, where $Q_{ijkl}$ and $\lambda_{ijkl}$ are electrostrictive and magnetostrictive coupling coefficients.

In the thermodynamic analysis, the polarization is uniform and oriented along the $(\overline{111})_{pc}$ direction, and a harmonic approximation is employed to describe the AFM order parameter in the spin cycloid(*6*) . The relative energies of the system for the spin cycloid with different propagation directions were calculated and plotted in Fig. 1**B**. In the case of an isotropic compressive strain, as observed in $BiFeO_3/SrTiO_3$, the thermodynamically preferred directions for the spin cycloid propagation are equivalent to the unstrained case described in ref.(*5*). Consequently, the ferroelectric monodomain will have three equivalent spin cycloids $\mathbf{k}_1$, $\mathbf{k}_2$, and $\mathbf{k}_3$ associated with it, all of which are equally likely to dominate.

The calculations were repeated for a system with an anisotropic compressive strain. Reproducing the epitaxial strain state observed in $BiFeO_3/NdGaO_3$, where the magnitude of the compressive strain is largest along the *c*-axis, the breaking of symmetry leads to the lifting of the degeneracy between the three propagation directions. The $\mathbf{k}_1$ and $\mathbf{k}_2$ directions are equivalent and are less stable when compared to the $\mathbf{k}_3$ direction, where the total free energy is lowest. The ferroelectric monodomain in this case is therefore also expected to host a single variant of the spin cycloid.

The constructed free-energy was used to conduct phase-field simulations of both strain cases for further validation. A brief outline of the simulation setup is presented in supplementary. The simulation results corroborate the thermodynamics analysis and experimental results as shown in Fig. 3.

6. STEM measurements

Cross-sectional TEM lamella samples of $(111)_{pc}$ $BiFeO_3$ film on $NdGaO_3$ $(011)_O$ substrates were prepared using a dual-beam focused ion beam system (Helios G3, FEI) for interfacial anisotropic strain analysis. A thin specimen was prepared using a Ga ion beam at 30 kV, with different acceleration voltages from 5 to 1 kV for sample cleaning to minimize Ga ion damage. STEM imaging was performed using a JEM-ARM200F (JEOL Ltd.) at the Materials Imaging & Analysis Center of POSTECH, equipped with a $5^{th}$ order aberration corrector (ASCOR, CEOS GmbH) to form a 0.7 Å probe. The accelerating voltage and convergent semi-angle of the beam were 200 kV and 28 mrad, respectively. The collection of semi-angles ranged from 54 to 216 mrad for high-angle annular dark-field (HAADF) imaging. Raw images were radial-difference filtered to remove background noise (Filters Lite, HREM Research Inc.). Lattice strain maps were obtained by conducting geometrical phase analysis (GPA) on high-resolution HAADF-STEM images using a GPA plug-in (HREM Research Inc.) implemented in Digital Micrograph (Gatan Inc.). Two-phase images were calculated by selecting two non-parallel reciprocal lattice vectors from the image's *Power Spectrum* to generate a 2-D strain map. The a-axis and c-axis angles were defined as 0° and 90°, respectively. The reference regions were defined in the $NdGaO_3$ $(011)_O$ substrate regions with known lattice parameters.

7. NV diamond microscopy

Antiferromagnetic domains at the surface of $BiFeO_3$ films were mapped by Nitrogen Vacancy (NV) scanning microscopy from Qnami Quantum Microscope, ProteusQ$^{TM}$. We used parabolic tapered Quantilever$^{TM}$ MX+ diamond tips for their excellent signal-to-noise ratio and photon collection efficiency, making them well-suited for detecting the very small stray fields present in $BiFeO_3$. The NV center, consisting of a nitrogen defect and a neighboring vacancy in negative charge state ($NV^-$), serves as a single-atom quantum sensor by utilizing its spin triplet state ($m_s$=0,±1). In our setup, an external magnet is used to break the degeneracy of the $m_s$=±1 states and a microwave (MW) source frequency is swept to obtain electron spin resonance (ESR) (between $m_s$=0 to $m_s$=±1 states) which is then detected optically by measuring photoluminescence (PL) intensity. As the NV diamond tip scans the sample surface, the local stray magnetic field (B) projected along the NV− axis shifts the ESR spectra enabling the tracking of the local magnetic contrast. Here we have performed imaging in two modes namely dual iso-B and full-B. In the former, PL is measured at two MW frequencies near the FWHM of the ESR spectra and its difference ($PL(\nu_2)$-$PL(\nu_1)$) is used to generate the real space magnetic contrast. In the latter, full ESR spectra is fitted at each point and strength of local magnetic field is estimated quantitatively.

8. Piezo-force microscopy (PFM)

Ferroelectric domains were mapped at room temperature using Piezo Force Microscopy (PFM) with a Jupiter XR Atomic Force Microscope from Oxford Instruments (Asylum Research). The Dual AC Resonance Tracking (DART) mode was used to record images by tracking the contact resonance frequency and adjusting the cantilever's drive frequency via a feedback loop. To write a domain using an electric field, the bottom $SrRuO_3$ layer was grounded, and the PFM tip applied the switching voltage from the top.

9. Switching of ferroelectric polarization under Au electrode for NV scans

For ferroelectric switching, a vertical capacitor structure of $BiFeO_3$ was fabricated with ~25 nm Au as the top electrode and ~30 nm $SrRuO_3$ as the bottom electrode. A similar triangular electric pulse with a frequency of 10 kHz and an amplitude of 400 kV/cm was used to switch the ferroelectric polarization via probe tip. Then, using an etchant (KI: $I_2$: $H_2O$= 4: 1: 40) Au electrode was removed. Subsequently, PFM was used to make sure that the polarization is switched by 180° and the scanning NV images on a switched region was carried out to probe the spin cycloid.

10. Non-local magnon transport

Transport measurements were conducted employing 4-terminal non-local devices, wherein two terminals were dedicated to source current injection, and the remaining two served as output terminals for inverse spin Hall effect (ISHE) voltage measurement. The entire experimental setup and procedures were controlled using an in-house developed Python code and a Keithley 7001 switch box, maximizing repeatability. See Supplementary Note 1 for details of the experimental protocols.

Using lock-in amplifiers, we separate higher-order contributions in the voltage by measuring higher harmonics, using: $V = R_1 I + R_2 I^2$, where $R_i$ is the $i^{th}$ harmonic response. We have used the thermal effects due to Joule heating (for which $\Delta T$ is proportional to $I^2$) are detected in the second harmonic signal. $R_2$ is the resistance of the second (2ω) harmonic response. This comprehensive setup allowed us to perform accurate and controlled transport measurements (using all automated codes), facilitating the investigation of magnon transport in the antiferromagnetic insulator. All the measurements were recorded in the absence of the magnetic/electric field, thus eliminating any interference during the inverse spin Hall voltage measurement.

**Supplementary Text 2**

**Phase-field Model and Simulations**.

The total free energy of an inhomogeneous system under an applied electric field '**E**' considering order parameters for polarization '**P**', antiferromagnetic order (AFM) '**L**', oxygen octahedral tilt (OOT) '**θ**', their gradients, and coupling terms between the orders in $BiFeO_3$ is given by

$$F = \int dV \Big\{ \alpha_{ij} P_i P_j + \alpha_{ijkl} P_i P_j P_k P_l + K_1 (L_1^2 L_2^2 + L_2^2 L_3^2 + L_3^2 L_1^2) + K_2 L_1^2 L_2^2 L_3^2 + h(\mathbf{L} \cdot \mathbf{P})^2 + \frac{1}{2} g_{ijkl} \frac{\partial P_i}{\partial x_j} \frac{\partial P_k}{\partial x_l} + A \sum_{i=1}^{3} (\nabla L_i)^2 + \gamma \mathbf{P} \cdot [\mathbf{L}(\boldsymbol{\nabla} \cdot \mathbf{L}) - (\mathbf{L} \cdot \boldsymbol{\nabla})\mathrm{L}] + \frac{1}{2} c_{ijkl} (\varepsilon_{ij} - \varepsilon_{ij}^0)(\varepsilon_{kl} - \varepsilon_{kl}^0) \Big\},$$

where $g_{ijkl}$ is the gradient energy coefficient tensor for polarization, and $\kappa_{ijkl}$ is the gradient energy coefficient tensor for oxygen octahedral tilt. The evolution of polarization and the OOT are governed by the following equations(*5*, *7*).

$$\frac{\partial \boldsymbol{P}}{\partial t} = -K_P \frac{\delta F}{\delta \boldsymbol{P}}$$

$$\frac{\partial \boldsymbol{\theta}}{\partial t} = -K_\theta \frac{\delta F}{\delta \boldsymbol{\theta}}$$

The governing equation for the evolution of the AFM order parameter is

$$\frac{\partial \boldsymbol{L}}{\partial t} = -\frac{\omega}{\upsilon \mu_0 L_S} \frac{\delta F}{\delta \boldsymbol{L}}$$

The semi-implicit Fourier spectral method is used to numerically solve the governing equations. To curtail the effects of periodic boundary conditions on the spin cycloids, buffer layers are applied at the boundaries of the system while solving **L.**

The stress-free strain is calculated by considering the contributions from electrostriction and magnetostriction

$$\varepsilon_{ij}^0 = Q_{ijkl} P_k P_l + \kappa_{ijkl} \theta_k \theta_l + \lambda_{ijkl} L_k L_l$$

The total strain $\varepsilon_{ij}$ is obtained by solving the mechanical equilibrium equation

$$\nabla \cdot \sigma = 0,$$

which is solved in the Fourier space with periodic boundary conditions or with thin-film boundary conditions.

**Supplementary Text 3**

The ferroelectric domain structures in Fig. **1D** of the manuscript were obtained by simulating the electrical switching of a pristine ferroelectric domain using an AFM tip. The applied voltage ($V_{\text{Applied}}$ = -4V) induces the nucleation of switched domains and subsequent growth. A computational domain of size 96×96×192 is used (Δx=Δy=Δz=0.5 nm), where the thickness of the film, $t_{Film}$ = 60 nm, and the thickness of the substrate, $t_{Substrate}$= 32 nm. The misfit strain ($\varepsilon_{xx} = -2.8\%, \varepsilon_{yy} = -2.3\%, \varepsilon_{xy} = 0.0\%$) is selected to match the experimentally obtained values for BFO on $NdGaO_3$ $(011)_O$, and ($\varepsilon_{xx} = -2.4\%, \varepsilon_{yy} = 1.1\%, \varepsilon_{xy} = 0.0\%$) is the same in the case of BFO on $DyScO_3$ $(011)_O$. Since the AFM order does not have a major influence on the electrical switching of the ferroelectric domain, it is neglected for this calculation.

**Supplementary Text 4**

The AFM domain patterns in Figs. **3F** of the main-text is obtained by evolving polarization along the in-plane direction ($[-1\text{-}1\text{-}1]_{pc}$), and the AFM order parameter is initialized using randomly-aligned spins. The computational domain is discretized using a uniform 1024×1024×1 grid with grid spacing 1 nm, and the system evolved for 720,000-time steps (Δt = 0.0796 ns). The misfit strain ($\varepsilon_{xx} = -2.8\%, \varepsilon_{yy} = -2.3\%, \varepsilon_{xy} = 0.0\%$) is selected to match the experimentally obtained values for BFO on $NdGaO_3$ $(011)_O$. The values of the coefficients used in the simulations are listed in Tables S1, S2, and S3.

**Table S1: Coefficients of $BiFeO_3$ used in the free-energy functional expression**(*5*, *7*)**.**

| Coefficient | Value (in SI units) |
|---|---|
| $\alpha_{11}$ | $-3.580\times10^{8}$ $C^{-2}\cdot m^{2}\cdot N$ |
| $\alpha_{1111}$ | $3.000\times10^{8}$ $C^{-4}\cdot m^{6}\cdot N$ |
| $\alpha_{1122}$ | $1.188\times10^{8}$ $C^{-4}\cdot m^{6}\cdot N$ |
| $K_1$ | $-1.0\times10^{4}$ $J\cdot m^{-3}$ |
| $K_2$ | $3.1\times10^{4}$ $J\cdot m^{-3}$ |
| h | $-3.2\times10^{4}$ $C^{-2}\cdot m^{2}\cdot N$ |
| $g_{1111}$ | $4.335\times10^{-11}$ $C^{-2}\cdot m^{4}\cdot N$ |
| $g_{1122}$ | $-3.400\times10^{-12}$ $C^{-2}\cdot m^{4}\cdot N$ |
| $g_{1212}$ | $3.400\times10^{-12}$ $C^{-2}\cdot m^{4}\cdot N$ |
| A | $4.0\times10^{-12}$ $J\cdot m^{-1}$ |
| γ | $8.1\times10^{-4}$ $J\cdot C^{-1}$ |

**Table S2: Elastic stiffness, electrostrictive, and magnetostrictive coefficients using Voigt notation**(*5*, *7*)**. The magnetostrictive coefficients were obtained via fitting to the DFT results**(*8*) **from the literature.**

| Coefficient | Value (in SI units) |
|---|---|
| $C_{11}$ | $1.7794\times10^{11}$ Pa |
| $C_{12}$ | $0.9635\times10^{11}$ Pa |
| $C_{44}$ | $1.2200\times10^{11}$ Pa |
| $Q_{11}$ | $0.071$ $C^{-2}\cdot m^{4}$ |
| $Q_{12}$ | $-0.03$ $C^{-2}\cdot m^{4}$ |
| $Q_{44}$ | $0.02015$ $C^{-2}\cdot m^{4}$ |
| $\lambda_{11}$ | $-5.000\times10^{-6}$ |
| $\lambda_{12}$ | $2.500\times10^{-6}$ |
| $\lambda_{44}$ | $2.000\times10^{-6}$ |

**Table S3: Constants used in the evolution equation for the AFM order parameter L(*5*).**

| Constant | Value |
|---|---|
| Electron gyromagnetic ratio, $\omega$ | $2.21\times10^5$ m·A$^{-1}$·s$^{-1}$ |
| Gilbert damping ratio, $\upsilon$ | 0.5 |
| Saturation AFM magnetization, $L_S$ | $5.6\times10^5$ A·m$^{-1}$ |

**Supplementary Text 5**

**Strain calculation of BFO films grown on different $(011)_O$ substrates from synchrotron x-ray diffraction.**

To maintain the continuity of the lattice structure when there is a mismatch between the lattice parameters of the film and substrate, the crystal structure of the film is known to epitaxially strain, so that the in-plane lattice parameters of the film match the in-plane lattice parameters of the substrate.

The epitaxial strain is defined by Equation 1, where $\varepsilon_\zeta$ is the epitaxial strain along the crystallographic direction $\zeta$, $\zeta_s$ is the lattice parameter of the substrate along direction $\zeta$ and $\zeta_f$ is the lattice parameter of the film along direction $\zeta$.

$$\varepsilon_\zeta = \frac{\zeta_s - \zeta_f}{\zeta_f}$$

To calculate the epitaxial strain the unit cell of the orthorhombic substrate was transformed to a lower symmetry monoclinic unit cell, and the unit cell of the hexagonal film was transformed to an orthorhombic one, so that the *a*, *b*, and *c* axes of the substrate are collinear with the *a*, *b*, and *c* axes of the film respectively. The matrix transformation for the substrate and for the film are given below, by Equations. 2 and 3 respectively.

$$\begin{pmatrix} a \\ b \\ c \end{pmatrix}_{substrate_m} = \begin{pmatrix} 1 & 0 & 0 \\ 0 & 1 & -1 \\ 0 & 2 & 1 \end{pmatrix} \begin{pmatrix} a \\ b \\ c \end{pmatrix}_o$$

$$\begin{pmatrix} a \\ b \\ c \end{pmatrix}_{film_o} = \begin{pmatrix} 1 & 0 & 0 \\ 1 & 2 & 0 \\ 0 & 0 & 1 \end{pmatrix} \begin{pmatrix} a \\ b \\ c \end{pmatrix}_h$$

The orientation of the film, with respect to the substrate orientation for the various crystallographic symmetries, is shown below in Fig. S1.

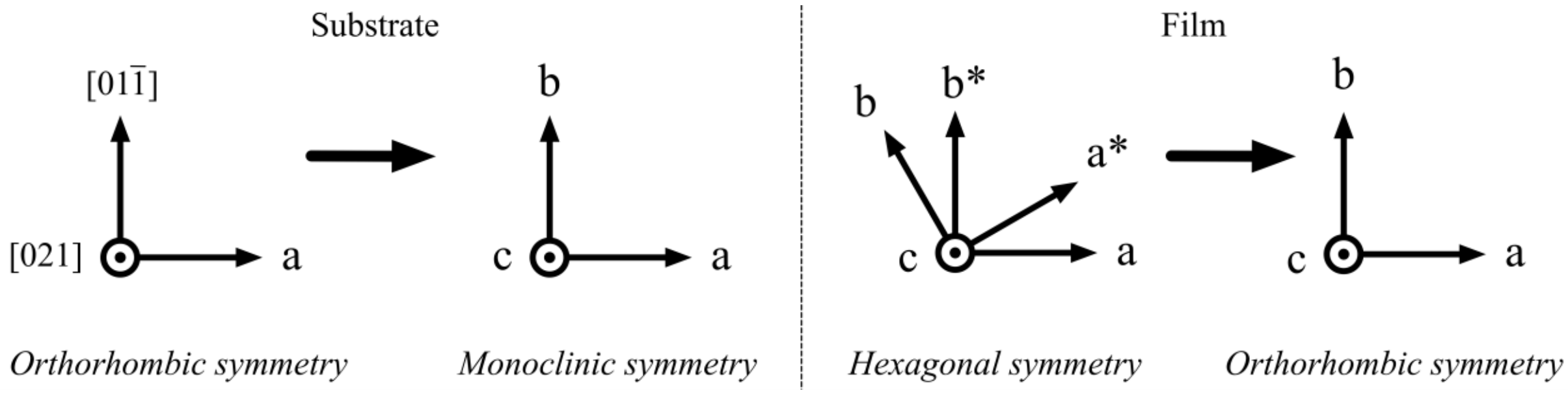

**Fig. S1. Conversion of various crystallographic symmetries.** The orientation of the film with respect to the substrate is shown for various crystallographic symmetries. The conventional crystallographic vectors are shown on the left and the transformed crystallographic vectors on the right.

The lattice parameters of the $NdGaO_3$ (NGO) substrate were averaged from Ref.(*9*) and Ref.(*10*), and the lattice parameters of TbScO3 (TSO) were taken from Ref.(*11*, *12*). The lattice parameters of the $BiFeO_3$ films grown on the NGO and TSO substrates were refined against the non-resonant x-ray scattering (NRXS) data in the hexagonal basis, but with no constraints and hence the label 'pseudo-hexagonal' is used to refer to the symmetry of the film. The lattice parameters of the various films and substrates are all tabulated in Table S4 together with the lattice parameters of the substrate in the monoclinic symmetry and the film in the orthorhombic symmetry. The epitaxial strain, given in Table S5, was calculated using Eqn. 1 using the in-plane lattice parameters of the film in the orthorhombic symmetry and the substrate in the monoclinic symmetry.

**Table S4. Lattice parameters are given for the $BiFeO_3$ film and the $NdGaO_3$, and $TbScO_3$ substrates in the conventional unit cell as listed in the ICSD database. The transformed unit cell lattice parameters are also listed for the film and substrates.**

| Symmetry | *a* (Å) | *b* (Å) | *c* (Å) | α (deg) | β (deg) | γ (deg) |
|---|---|---|---|---|---|---|
| $NdGaO_3$ Substrate | | | | | | |
| Orthorhombic | 5.430(3) | 5.500(3) | 7.712(3) | 90 | 90 | 90 |
| Monoclinic | 5.430(3) | 9.472(4) | 13.435(7) | 89.53(5) | 90 | 90 |
| 1000 nm $BiFeO_3$ grown on $NdGaO_3$ | | | | | | |
| Pseudo-hexagonal | 5.583(5) | 5.615(5) | 13.863(5) | 89.93(5) | 90.05(5) | 120.27(5) |
| Pseudo-orthorhombic | 5.583(5) | 9.70(3) | 13.863(5) | 89.95(5) | 90.05(5) | 90.5(3) |
| Bulk $BiFeO_3$ [6] | | | | | | |
| Hexagonal | 5.5804(2) | 5.5804(2) | 13.8721(7) | 90 | 90 | 120 |
| | 5.580(7) | 9.666(9) | 13.872(5) | 90 | 90 | 90 |

**Table S5. The epitaxial strain applied by the NGO and TSO substrates on the 1000 nm $BiFeO_3$ films along the *a* and *b* in-plane directions of the film in the orthorhombic symmetry and the substate in the monoclinic symmetry.**

| Substrate | Thickness of $BiFeO_3$ film | $\varepsilon_a$ | $\varepsilon_b$ |
|---|---|---|---|
| $NdGaO_3$ | 1000 nm | - 2.738(3) % | - 2.35(1) % |

The lattice parameters of the $BiFeO_3$ films are comparable to those of the bulk, which are also tabulated in Table 1, indicating the films are relaxed overall. Hence, it is valid to use these lattice parameters to evaluate the epitaxial strain. Even though the films are fully relaxed, the impact of the epitaxial strain can be observed in the domain formation of the films as detailed in the main paper.

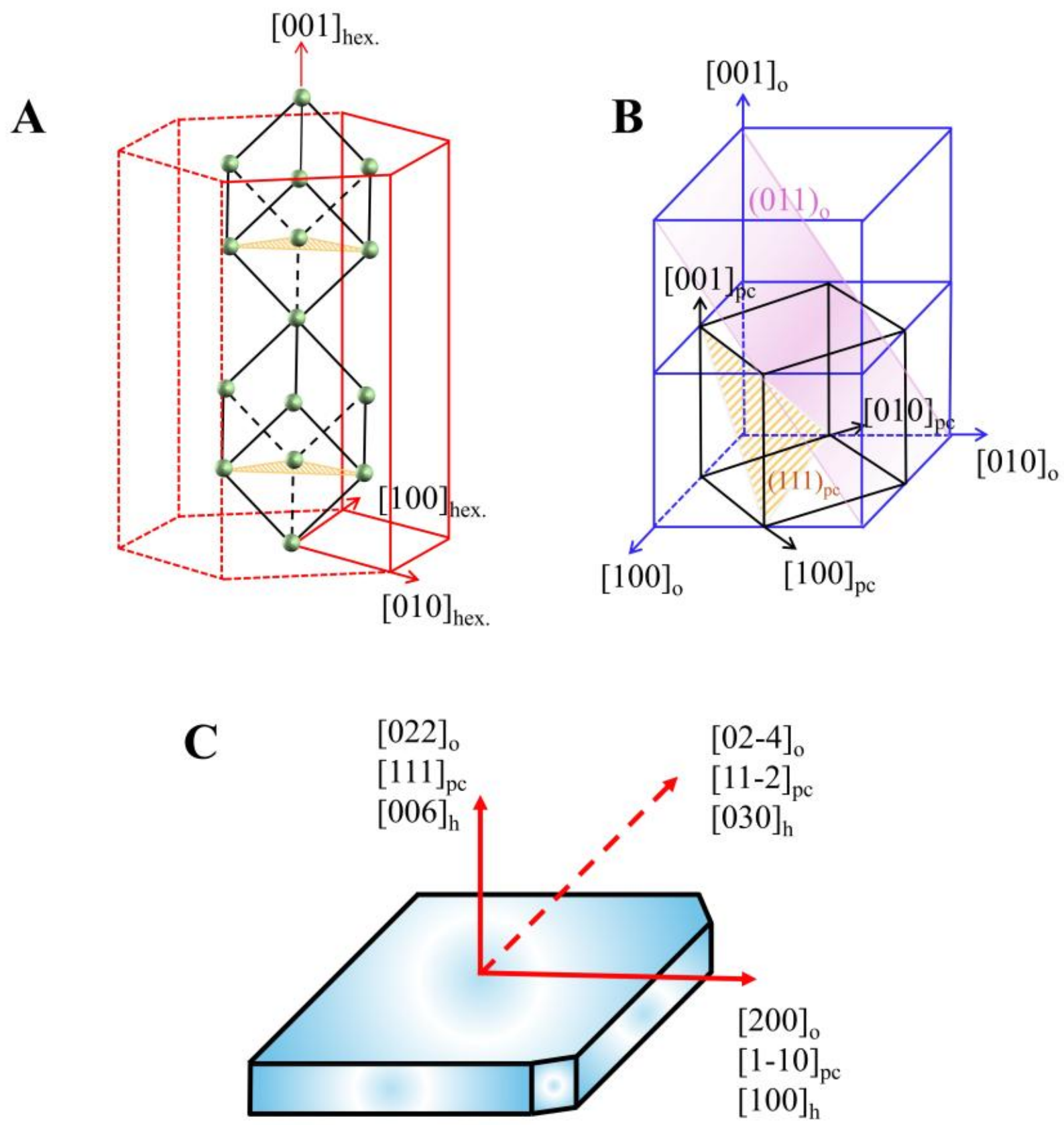

**Fig. S2. Crystallographic relationships between hexagonal, cubic and orthorhombic representations.** **(A)** Schematic representation of the crystallographic relationship between the hexagonal (red) and pseudocubic (black) $BiFeO_3$ unit cells. **(B)** Similar relationship between the orthorhombic (blue) and the pseudocubic (black) unit cells. **(C)** Relationship between crystallographic axes of different structural representations on the thin film used under this study.

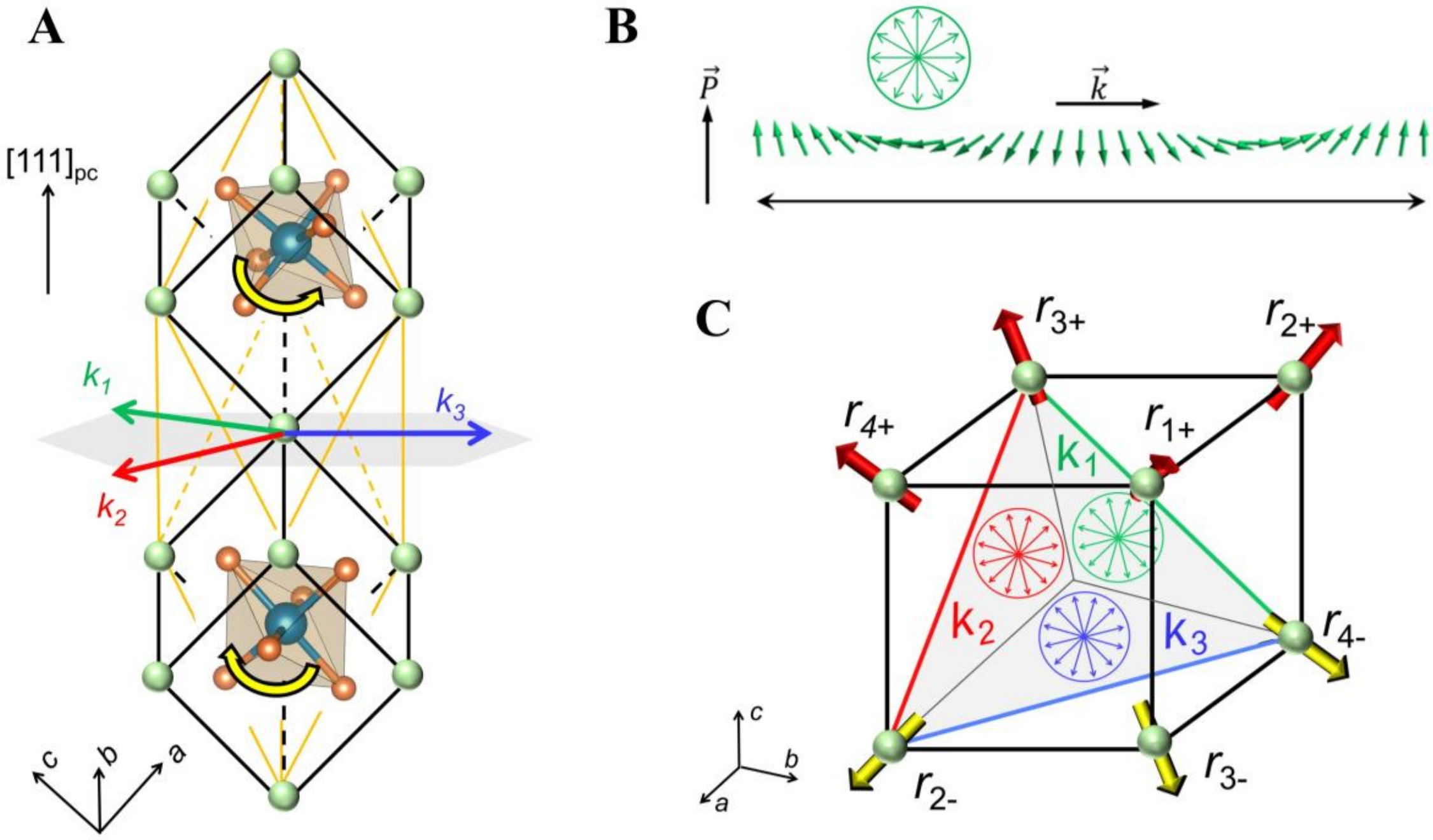

**Fig. S3. Complex multiferroic-domain population of $BiFeO_3$. (A)** Rhombohedral unit cell (yellow color) of $BiFeO_3$ with distortion along (111) and corresponding pseudocubic cell (black color). (**B)** A description of the type I spin cycloid of $BiFeO_3$, which is within the plane of polarization (**P**) and propagation vector (k). (**C)** Schematic illustration of the complex ferroelastic, ferroelectric, and antiferromagnetic domain (type I cycloid) population of $BiFeO_3$. Type I spin cycloid is with propagation vectors $k_1$ = $[-1,1,0]_{pc}$, $k_2$ =$[0,1,-1]_{pc}$, and $k_3$ = $[1,0,-1]_{pc}$.

Bulk $BiFeO_3$ exhibits a distorted rhombohedral structure at room temperature, usually represented as a pseudocubic unit cell(*13*) as shown in Fig. S1a. The ferroelectric polarization of $BiFeO_3$ arises due to the displacement of Bi ions along $<111>_{pc}$ direction. Whereas the antiferromagnetism comes from the Fe-O-Fe superexchange interaction, the non-collinear arrangement of spins is promoted by magnetoelectric interaction driven off-center distortion of Fe ions (*14*, *15*). Thus, due to the presence of competing superexchange (leading to collinear spin arrangement) and off-centric distortion (leading to non-collinearity) stabilizes incommensurate type-I spin-cycloid with the *k*-vector lying in the $(111)_{pc}$ plane along three equivalent <1-10> directions(*16*, *17*). Thus, on a larger scale, the overall net magnetic moments should be compensated, however, due to the presence of a relative rotation of $FeO_6$ octahedra along $[111]_{pc}$ direction, significant Dzyaloshinskii-Moriya interaction is induced which causes the spin-moment out-of-the plane of cycloid to result in a spin-density wave with the same period of ~65 nm(*18*).

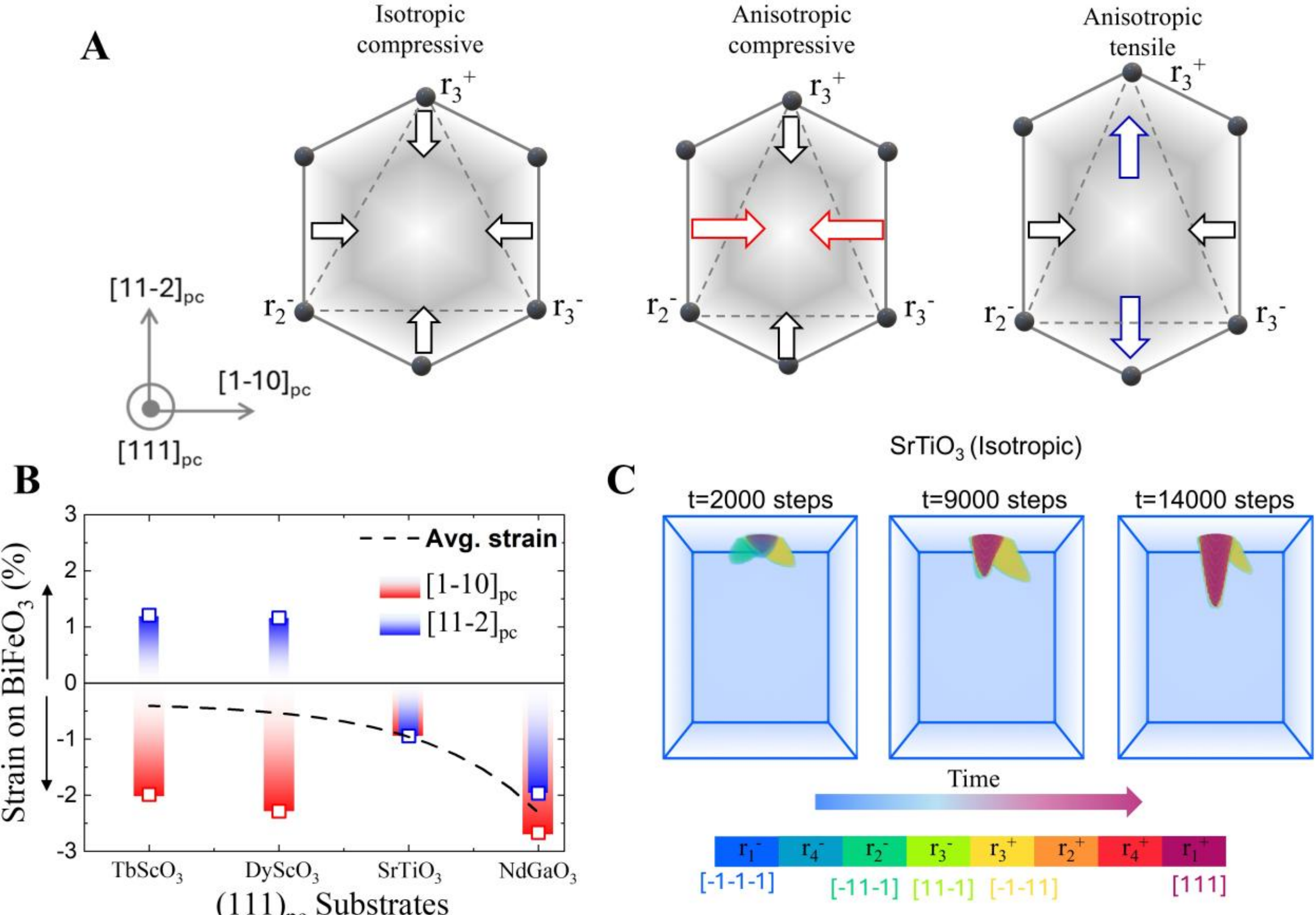

**Fig. S4. Effect of strain and symmetry**. **(A)** Schematic illustrations of the isotropic, anisotropic-compressive and anisotropic-tensile strains on the $(111)_{pc}$ plane. **(B)** Variation of in-plane strains along $[1\text{-}10]_{pc}$ and $[11\text{-}2]_{pc}$ on $(111)_{pc}$ BFO films grown at $TbScO_3$ (TSO), $DyScO_3$ (DSO), $SrTiO_3$ (STO), and $NdGaO_3$ (NGO) substrates in their $(011)_O$ configurations. **(C)** Ferroelectric switching process calculated via phase field simulations of (111) BFO thin-film grown on $SrTiO_3$ with isotropic-compressive strain. Minority domains are formed due to intermediate 71° switching, consistent with previous calculations, which is primarily responsible for polarization fatigue(*19*). Similar data for $REScO_3$ (RE=Tb or Dy) and $NdGaO_3$ are shown in Fig. 1**D**.

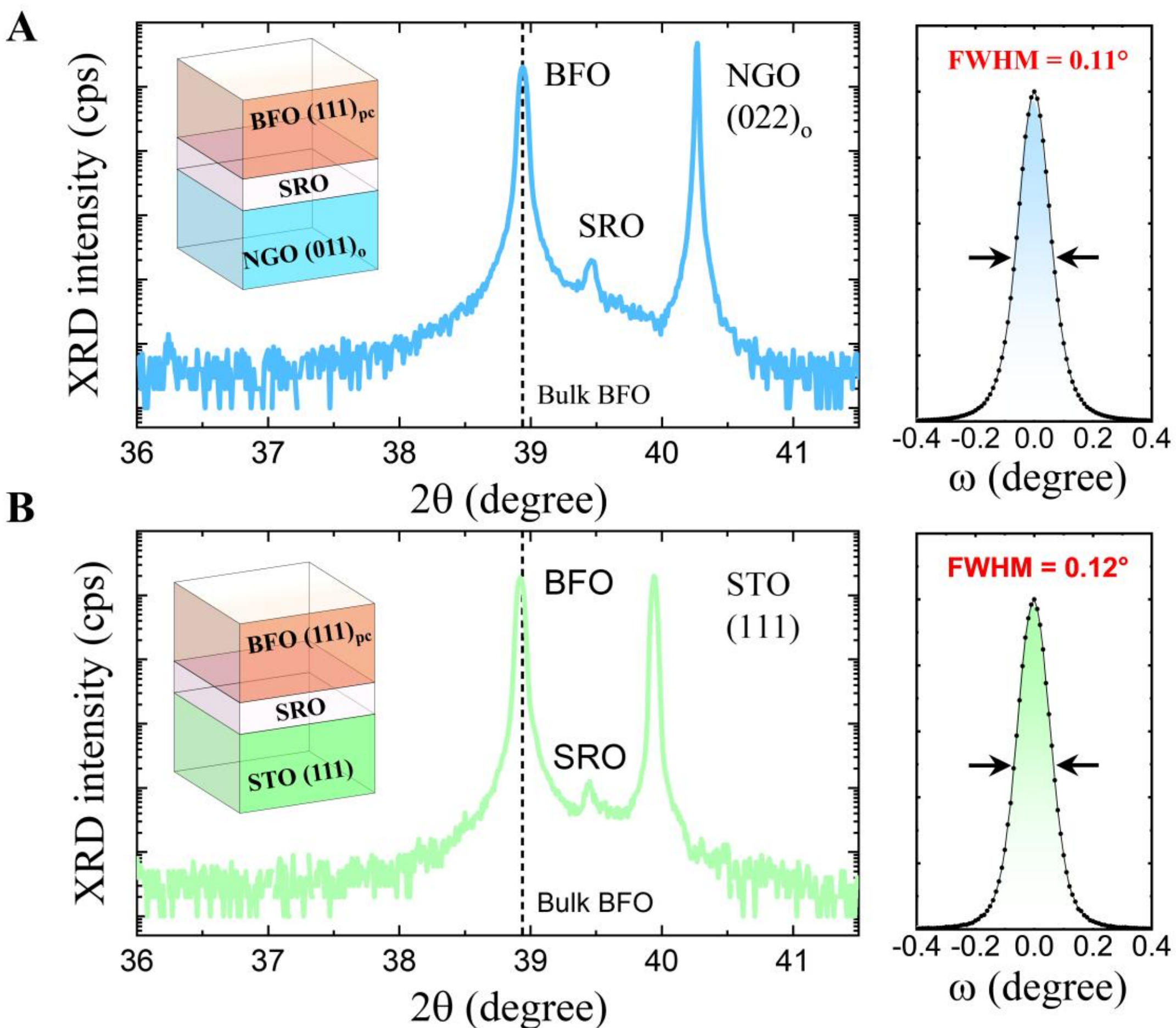

**Fig. S5. Structural characterization of (111) $BiFeO_3$ films.** Room temperature XRD spectra (along with corresponding rocking curves around (111) peak) of $BiFeO_3$ (BFO) thin films grown on **(A)** orthorhombic $(011)_O$ or $(111)_{pc}$ $NdGaO_3$ (NGO), and **(A)** (111) $SrTiO_3$ (STO) buffered by ~25 nm $SrRuO_3$ (SRO) layers. Rocking curve of (111) $BiFeO_3$ peak with full-width-at-half-maximum (FWHM) of ~0.1° indicates good crystalline quality*(*20*)*.

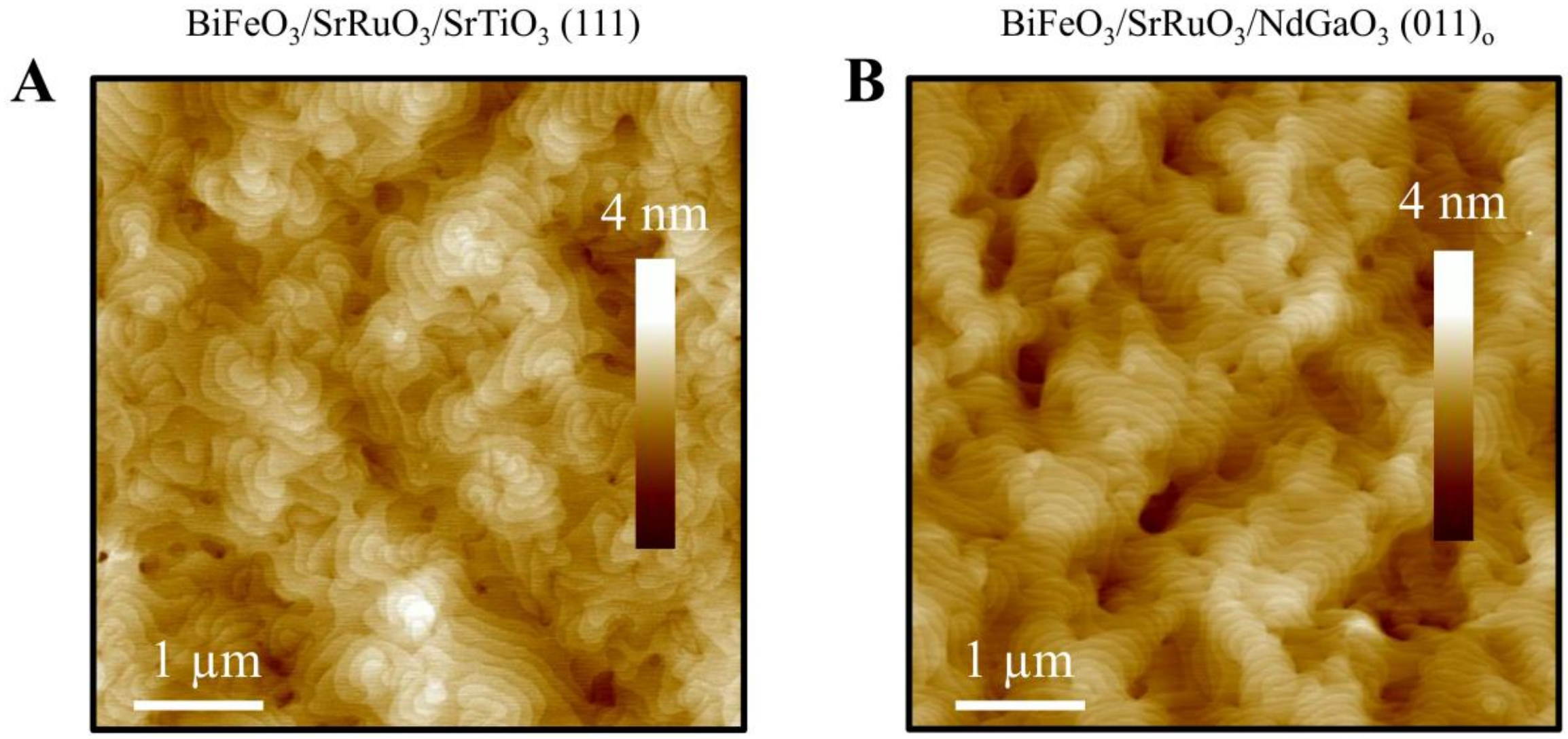

**Fig. S6. Atomic force microscopy images of $BiFeO_3$ films grown on $SrTiO_3$ and $NdGaO_3$.** Atomic force microscopy images of 1000 nm $BiFeO_3$ films grown on **(A)** $SrTiO_3$ and **(B)** $NdGaO_3$ display an atomically smooth surface with a root-mean-square (RMS) roughness of ~3Å (which is much better than previous reports(*20*) of (111) $BiFeO_3$ film), as desirable for magnetoelectric device fabrications.

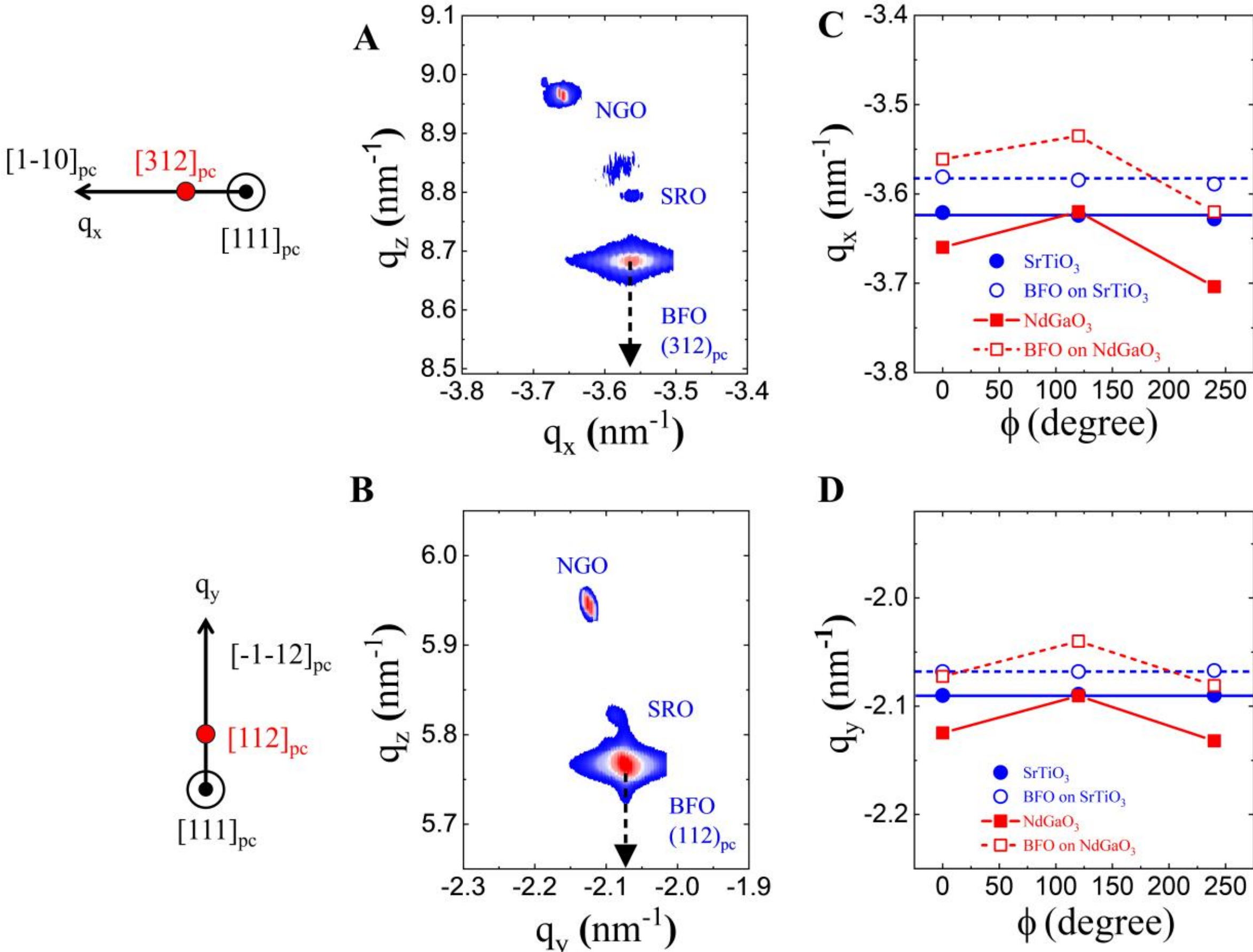

**Fig. S7. RSM spectra for BFO films on both $SrTiO_3$ and $NdGaO_3$ substrates. (A)** Reciprocal Space Mapping (RSM) spectra taken around in-plane $(312)_{pc}$ asymmetric peak which is within the plane consisting of $[111]_{pc}$ and $[1\text{-}10]_{pc}$. **(B)** Corresponding spectra for $(112)_{pc}$ peak which is within the plane consisting of $[111]_{pc}$ and $[\text{-}1\text{-}12]_{pc}$. **(C)** and **(D)** display variations of extracted $q_x$ and $q_y$ intercepts with the in-plane $\phi$ angle for $(312)_{pc}$ and $(112)_{pc}$ peaks respectively, which clearly show the evidence of the lifting of cubic 3-fold symmetry for (111) $BiFeO_3$ films grown on $NdGaO_3$ in comparison to that of $SrTiO_3$.

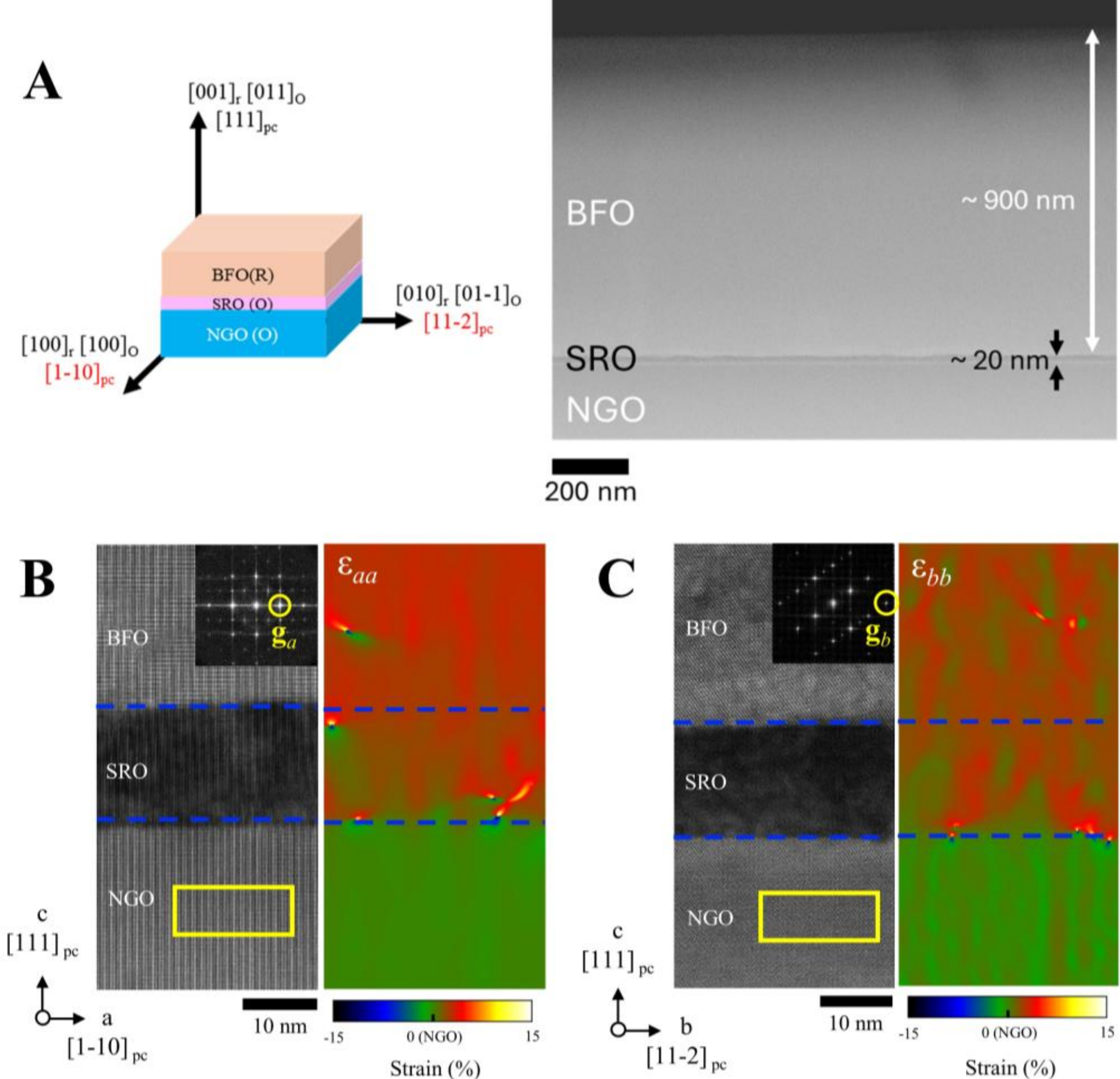

**Fig. S8. Scanning transmission electron microscopy of (111) $BiFeO_3$ film on $NdGaO_3$.** **(A)** Cross-sectional STEM image of $BiFeO_3$ (BFO) grown on $NdGaO_3$ (NGO) shows a defect-free, high-quality single domain. **(B-C)** HAADF images and in-plane strain maps along the B ($[11\text{-}2]_{pc}$) and A ($[1\text{-}10]_{pc}$) projections, respectively. Yellow circles indicate reciprocal lattice vectors ($\overrightarrow{\boldsymbol{g_x}}$) used for geometric phase analysis (GPA), with in-plane lattice strain ($\varepsilon_{xx}=(X - X_{NGO})/\ X_{NGO}$) calculated from HAADF-STEM. The reference regions for GPA were set on the $NdGaO_3$ substrates (yellow boxes in **(B)** and **(C)**).

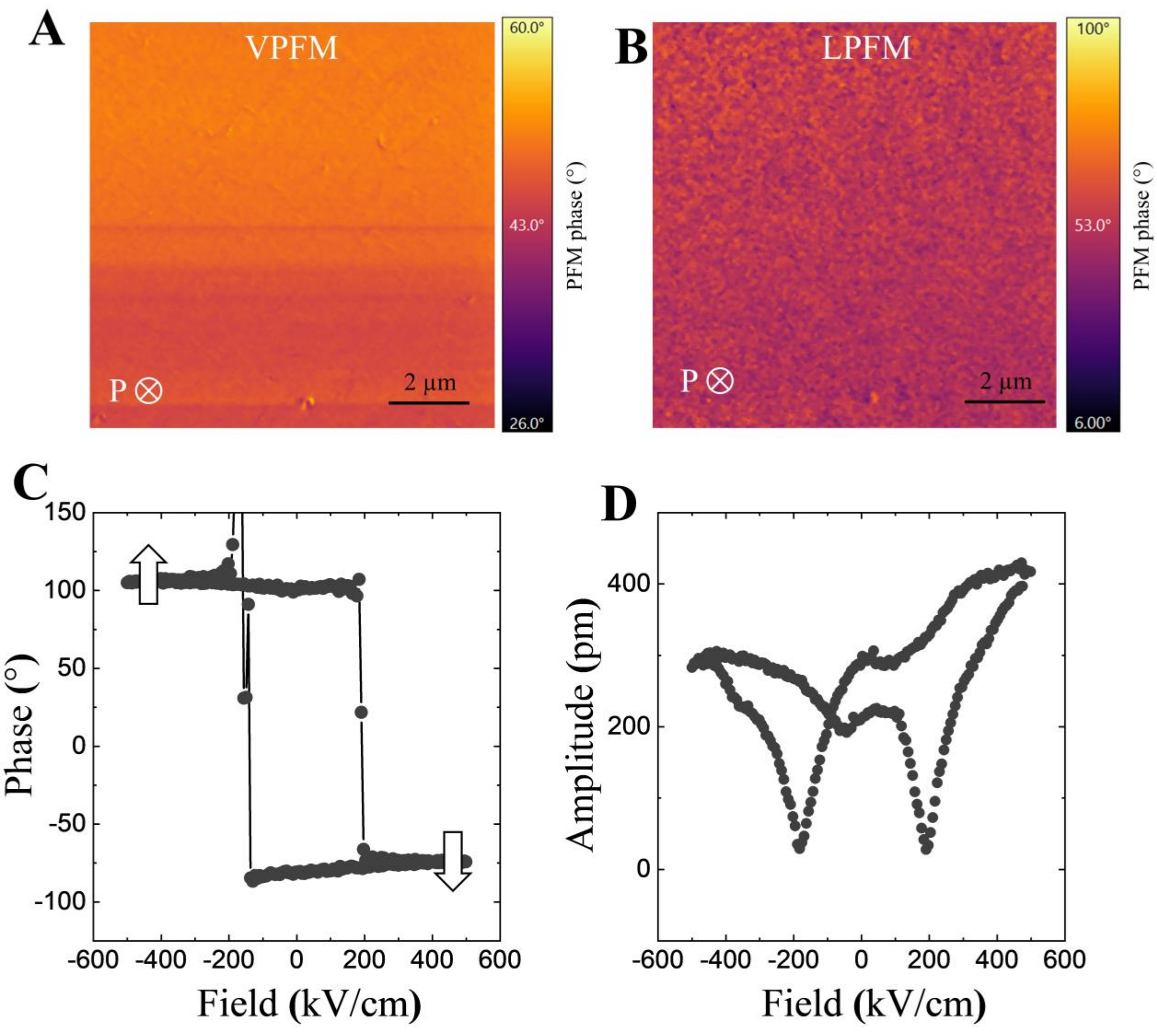

**Fig. S9. Piezoforce microscopy (PFM) of $(111)_{pc}$ BFO film on $NdGaO_3$. (A)** Vertical and **(B)** lateral PFM images display single ferroelectric domain nature of $BiFeO_3$ films grown on $NdGaO_3$ (011)o substrate. **(C)** Vertical piezoelectric hysteresis loop through local switching by PFM tips. **(D)** Variation of corresponding amplitude. This data clearly shows 180° polarization reversal. DC electric field was applied through the PFM tip, while the bottom $SrRuO_3$ layer was grounded.

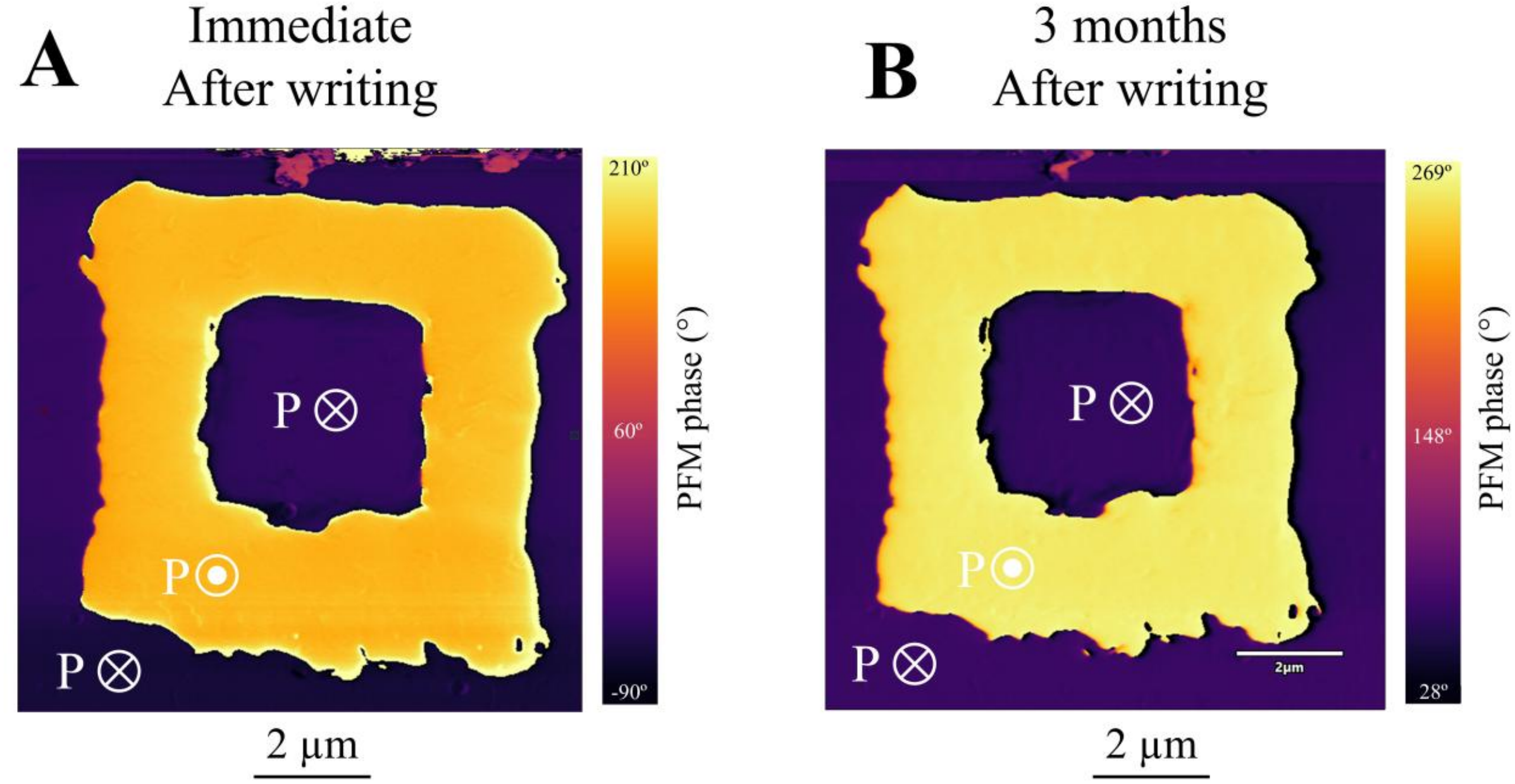

**Fig. S10. Robust nonvolatile ferroelectric writing in single domain $(111)_{pc}$ BFO.** Vertical PFM images for **(A)** immediately after the electric field (±50 kV/cm) writing using a PFM tip. **(B)** Corresponding reading of the same region after 3 months (2160 hours), strongly highlighting the non-volatile nature of the electric field writing in single domain $(111)_{pc}$ BFO grown on $NdGaO_3$.

**Discussions:** Such a permanent retention of the ferroelectric domain can be understood from the consideration of electrical boundary conditions. High asymmetry or deviation of ferroelectric PE loops can be caused due to different electrodes at bottom and top of the thin film respectively, which is primarily considered as the driving mechanism for the ferroelectric relaxation or back switching, as it causes strong depolarization field. Recently it was found that controlling this, a permanent ferroelectric retention was achieved in $BiFeO_3$ mesocrystal as discussed in ref.(*19*). In this regard, it is important to note that our $BiFeO_3$ films on $NdGaO_3$ exhibits a merely ~20% asymmetry in the PE loop (supplementary Fig. 6) which is almost four times smaller than what observed in $BiFeO_3$ mesocrystal(*19*). Additionally, the 180° ferroelectric switching in these $(111)_{pc}$ thin film does not lead to any ferroelastic domain deformation. Thus, there is no competition for elastic energy at the boundary of the unswitched and switched ferroelectric domains. Hence, these $BiFeO_3$ films on $NdGaO_3$ exhibit permanent ferroelectric retention, which is excellent and a key requirement for non-volatile magnetoelectric memory applications.

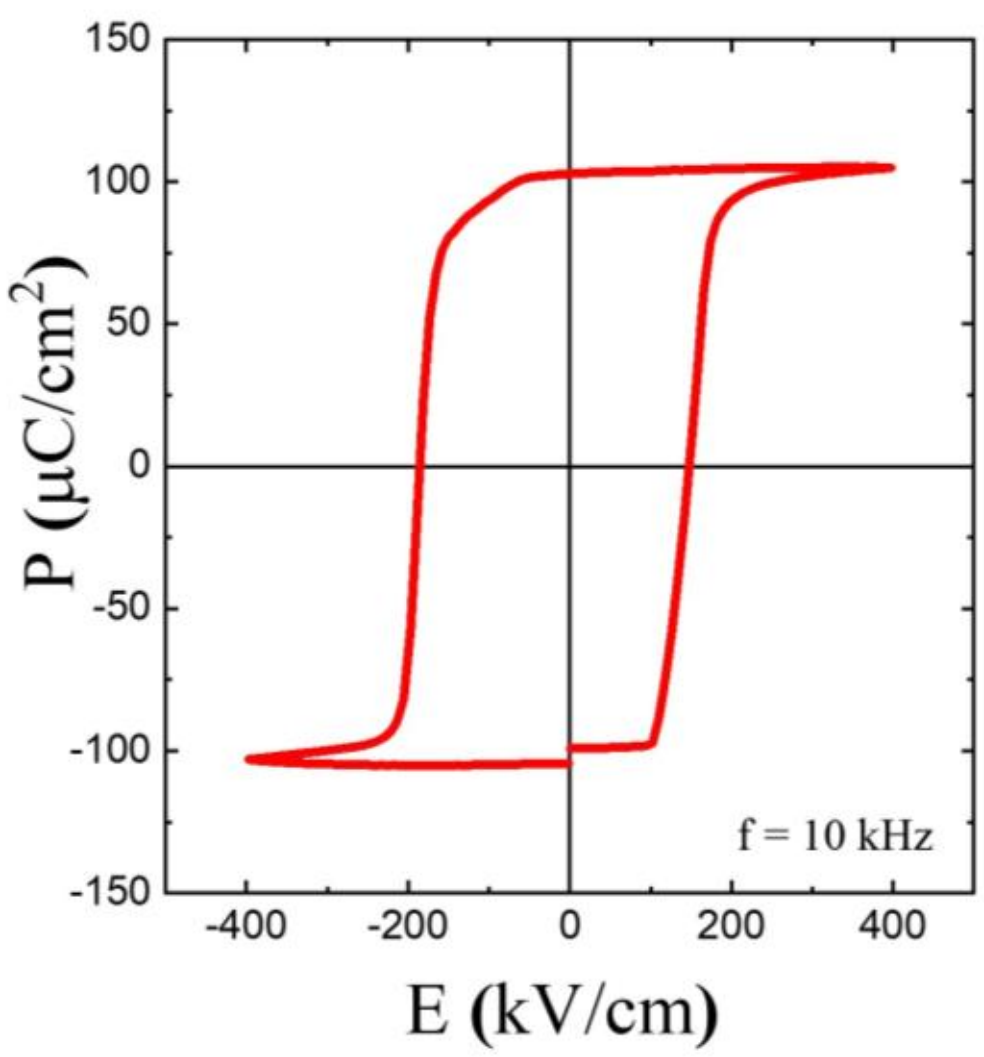

**Fig. S11. Ferroelectric polarization loop.** Ferroelectric polarization loop of a 1000 nm $(111)_{pc}$ $BiFeO_3$ thin film grown on $NdGaO_3$.

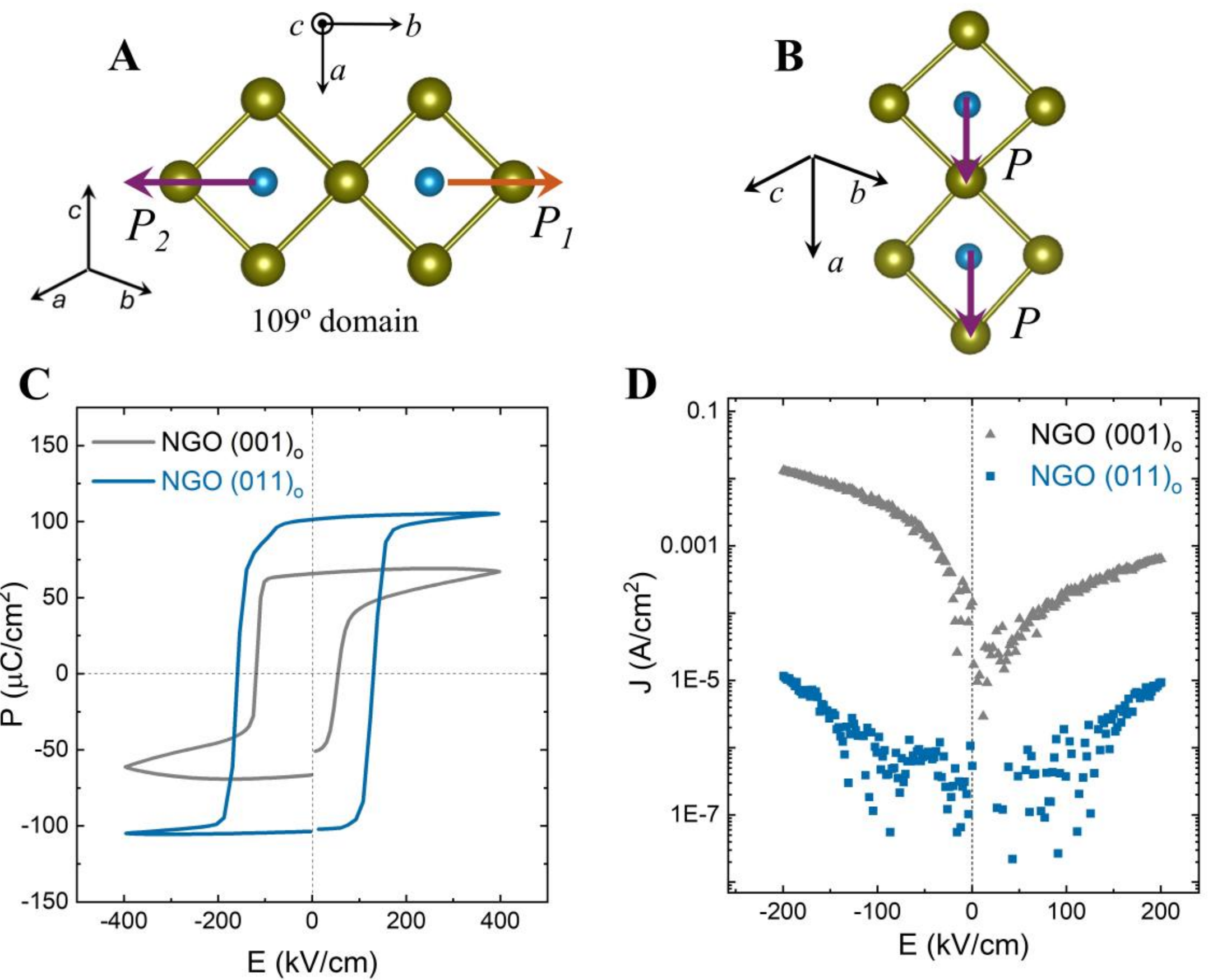

**Fig. S12. Leakage response of $BiFeO_3$ film grown on $NdGaO_3$ $(011)_O$ and $(001)_o$ substrates. (A-B)**, Schematic representation of the presence of 109º and single ferroelectric domains when $BiFeO_3$ (BFO) grown on $NdGaO_3$ (NGO) substrates with orientations $(001)_O$ and $(011)_O$, respectively. **(C-D)** Their corresponding P vs E loops, and J vs E leakage behavior.

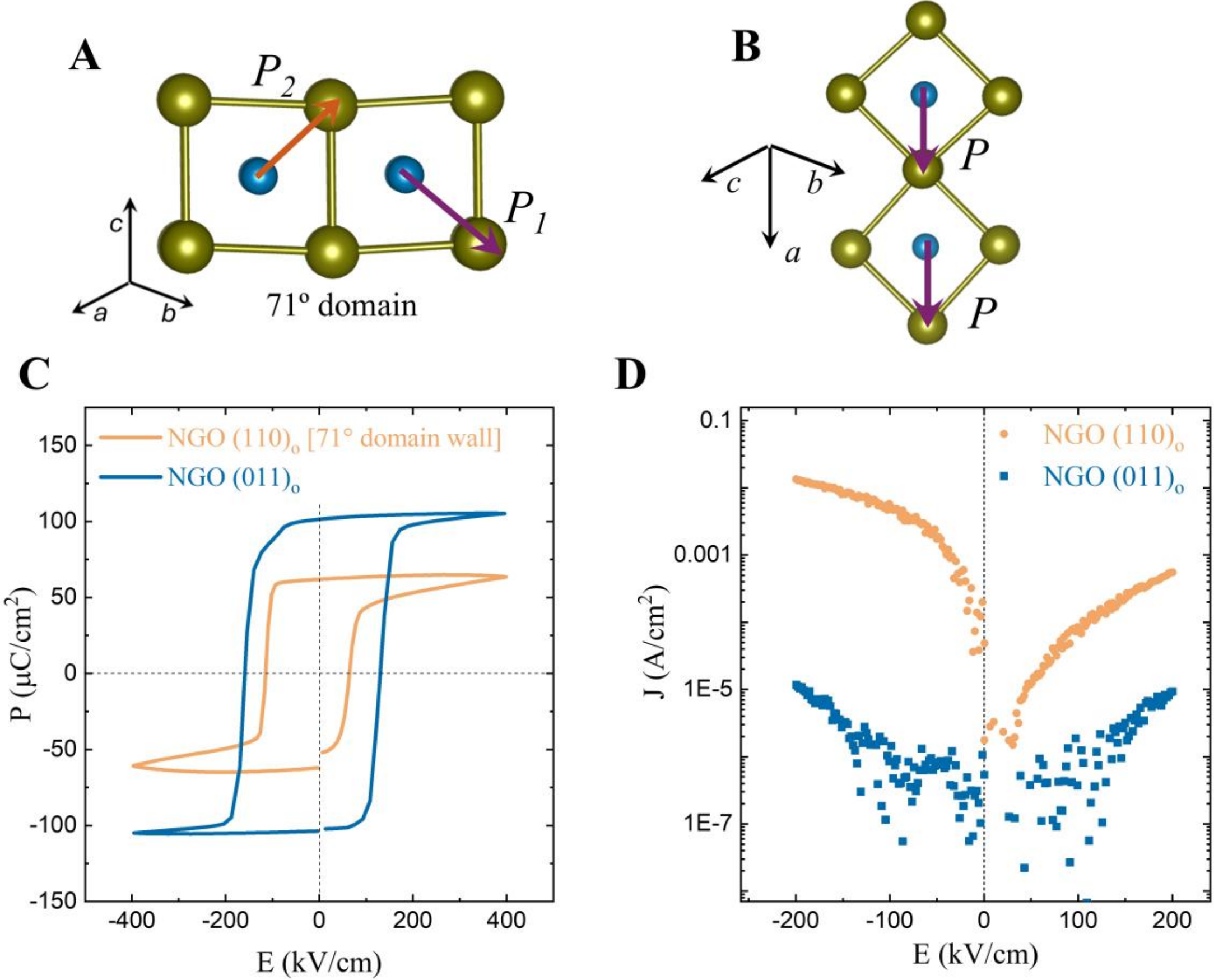

**Fig. S13. Leakage response of $BiFeO_3$ film grown on $NdGaO_3$ $(011)_O$ and $(110)_O$. (A-B)**, Schematic representation of the presence of 71º and single ferroelectric domains when $BiFeO_3$ (BFO) grown on $NdGaO_3$ (NGO) substrates with orientations $(011)_O$ and $(110)_O$ respectively. **(C-D)** Their corresponding P vs E loops, and J vs E leakage behavior.

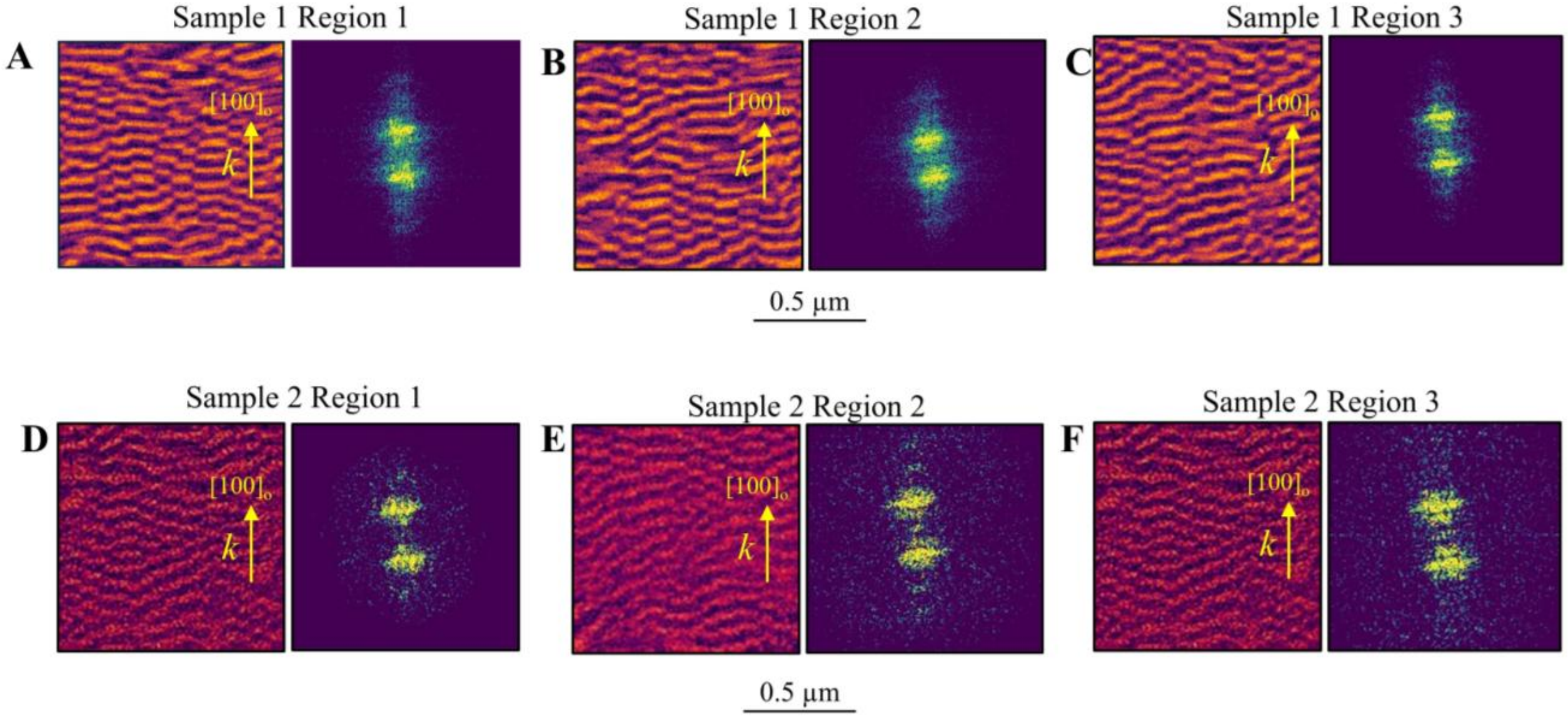

**Fig. S14. Location and sample independent robust single spin cycloid of $(111)_{pc}$ $BiFeO_3$ on $NdGaO_3$. (A-C)**, NV scanning and the corresponding FFT images at different locations of the sample-1 used in the main manuscript. **(D-E)**, NV scanning and the corresponding FFT image of the second sample at different samples. Importantly, all these scans indicate the presence of single spin-cycloid in $BiFeO_3$ on $NdGaO_3$ $(011)_O$.

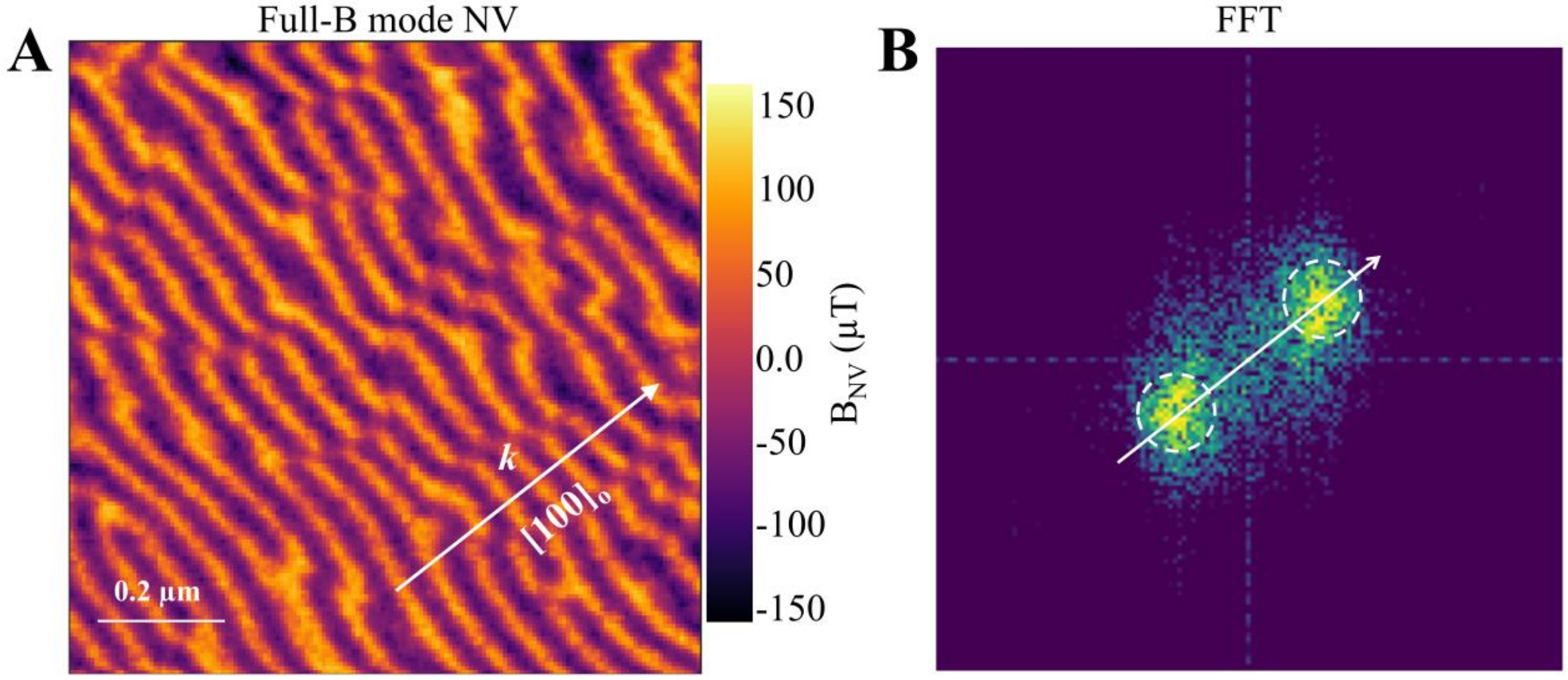

**Fig. S15. Scanning NV microscopy in full-B mode. (A)** Mapping of magnetic contrast collected in full-B mode(*21*) of scanning NV microscopy on $BiFeO_3/SrRuO_3/NdGaO_3$ $(011)_O$ thin film, which resembles the modulation of the actual stray magnetic field at the nanometer local scale. **(B)** Corresponding fast-Fourier-transform (FFT) image showing clear presence of single spin cycloid domain.

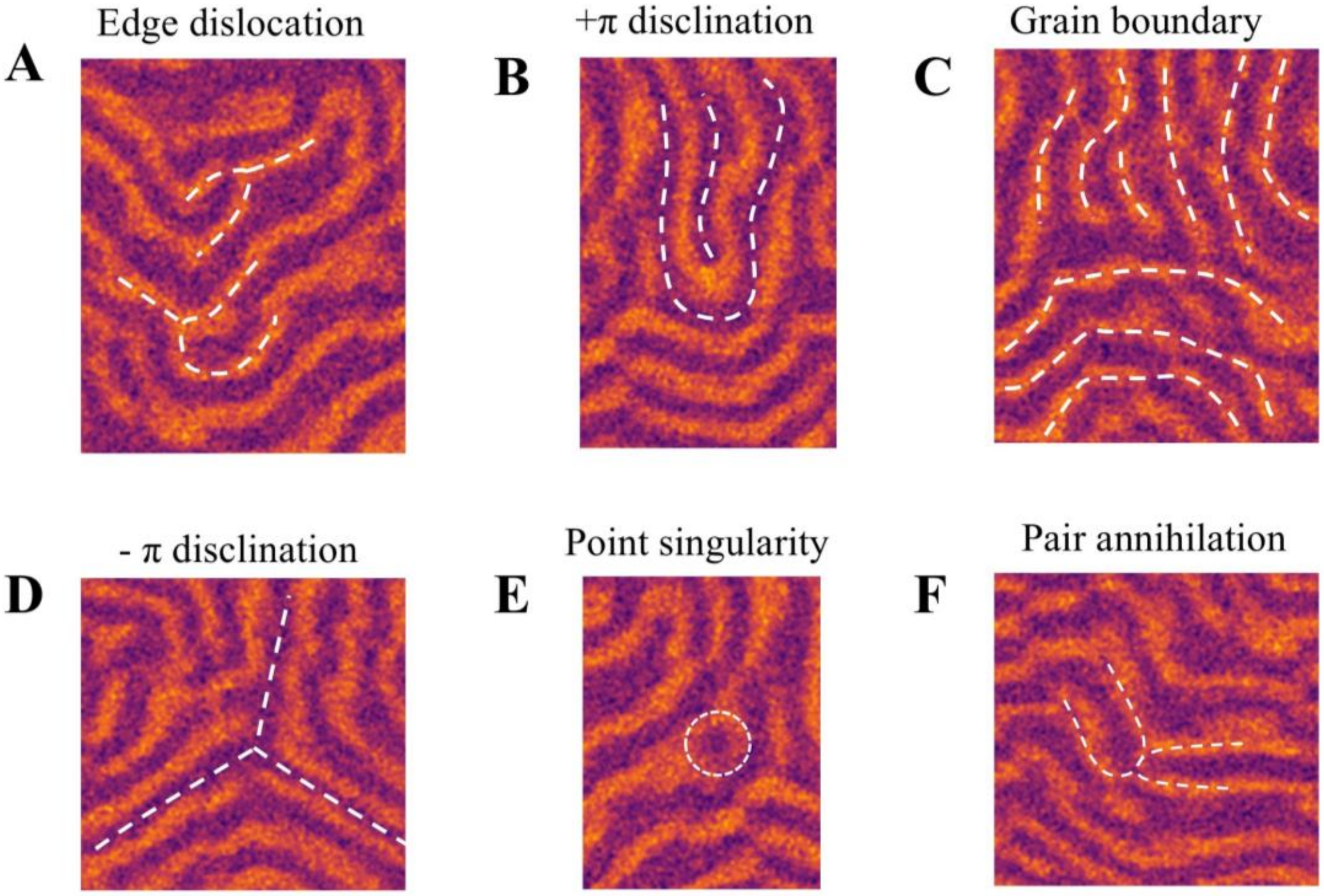

**Fig. S16. Spin-texture analysis of the scanning NV image of $BiFeO_3$ grown on (111) $SrTiO_3$.** Spin texture analyses show the presence of different topological defects e.g., **(A-F)** edge dislocation, +π disclination, grain boundary, -π disclination, point singularity and pair annihilation.

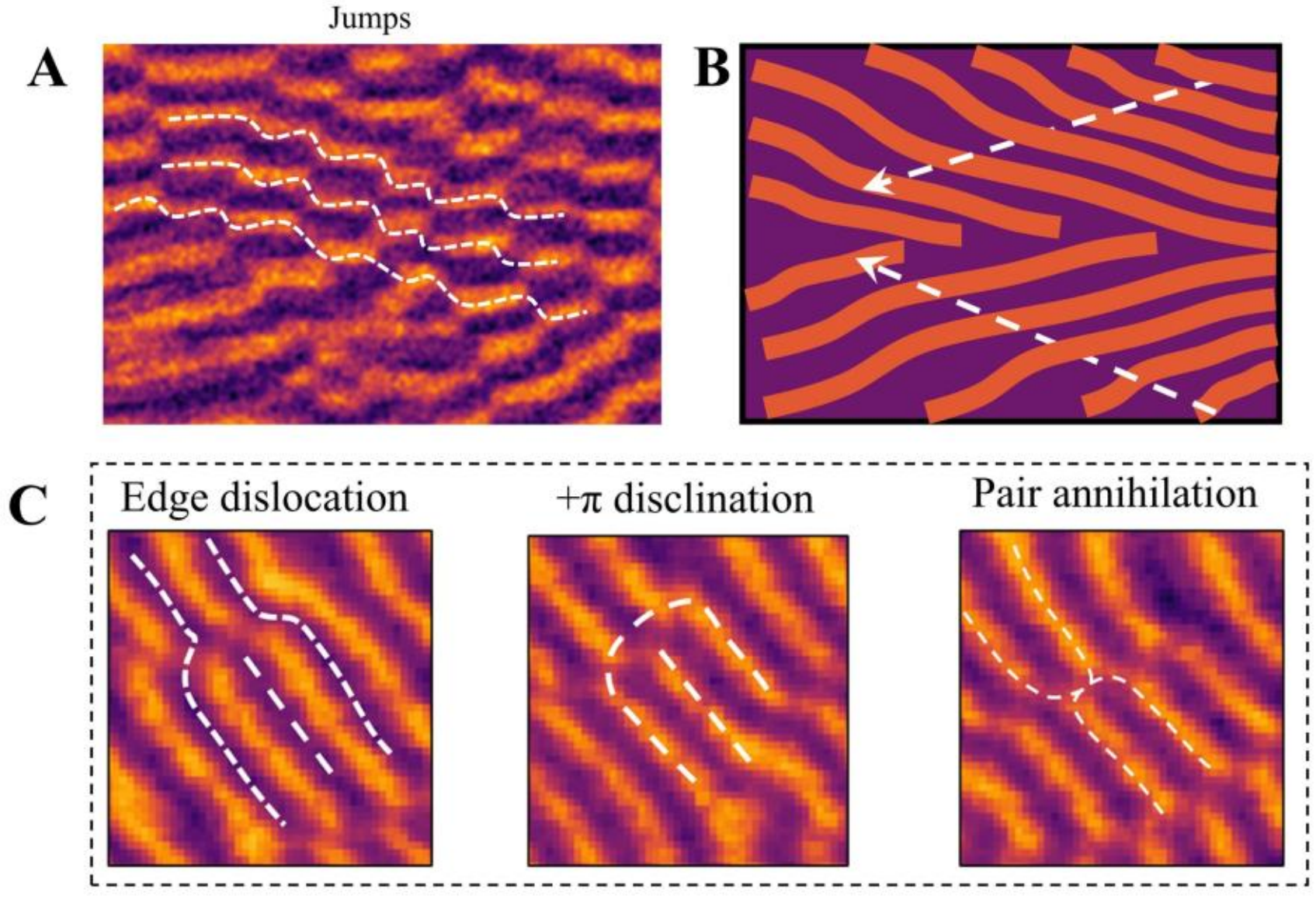

**Fig. S17. Spin-texture analysis of the scanning NV image of $BiFeO_3$ grown on $(011)_O$ $NdGaO_3$.** Spin texture analyses show the presence of different topological defects, e.g., **(A-E),** stochastic jumps, edge dislocations in pairs, edge dislocations, +π disclination, and pair annihilation.

**Supplementary Text 6**

**Spin texture analyses**

The spin texture of (111) $BiFeO_3$ on $SrTiO_3$ is of a complex pattern and has some intriguing features, which seem topological and possibly arise from some kind of instabilities present in both spatial and temporal domains. From the above figure, we note topological-line defects ±π disclinations and edge dislocations which are a consequence of different types of rotation/translation of the spin-cycloid propagation vector. The first arises due to the breaking of continuous rotation of the cycloid and the second one arises due to the breaking of translational symmetry(*22–24*). Importantly, features have a non-zero topological charge and can interact with each other which can be used to store and process information(*25*). There are additional topological point defects that can behave like vortex or antivortex of spin modulation leading to skyrmion(*26*). Further, some interactions can happen between these same/different (line and point) types of defects that can lead to hybrid topological defects as shown in the form of a hammer or pair annihilation or grain boundaries etc.(*24*, *27–29*).

The spin texture of $(111)_{pc}$ $BiFeO_3$ on $NdGaO_3$ is of a much simpler pattern than that on $SrTiO_3$, albeit with some intriguing features, which seem to have topological origin. From the above figure, we note the presence of only topological-line defects ±π disclinations and edge dislocations, primarily because of the lifting of antiferromagnetic domain degeneracy due to orthorhombic symmetry and anisotropic in-plane compressive strain by $NdGaO_3$ on $BiFeO_3$. Such a reduced symmetry breaks the continuous rotational symmetry and leads to the stripe-like pattern, which only gives rise to topological line defects due to spatial and temporal instability along a particular direction(*24*, *28*).

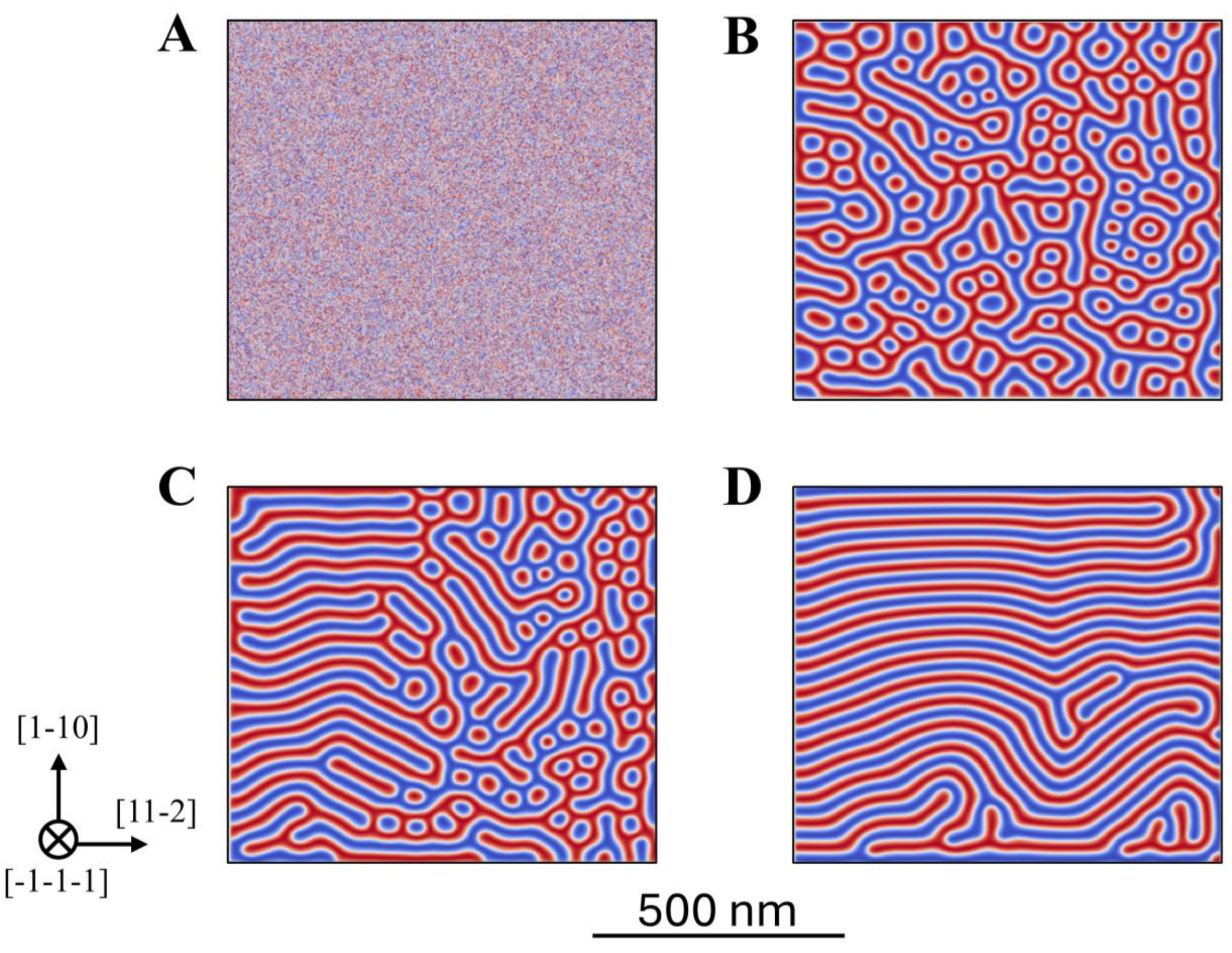

**Fig. S18. Evolution of the AFM order parameter in $BiFeO_3$ grown on the $(011)_O$ $NdGaO_3$ substrate. (A)** Initial configuration with random noises. **(B)** Domain pattern at t = 50,000-time steps, **(C)** 250,000 time steps and **(D)** 720,000 time steps. The colors represent the AFM component along out of the plane of the page.

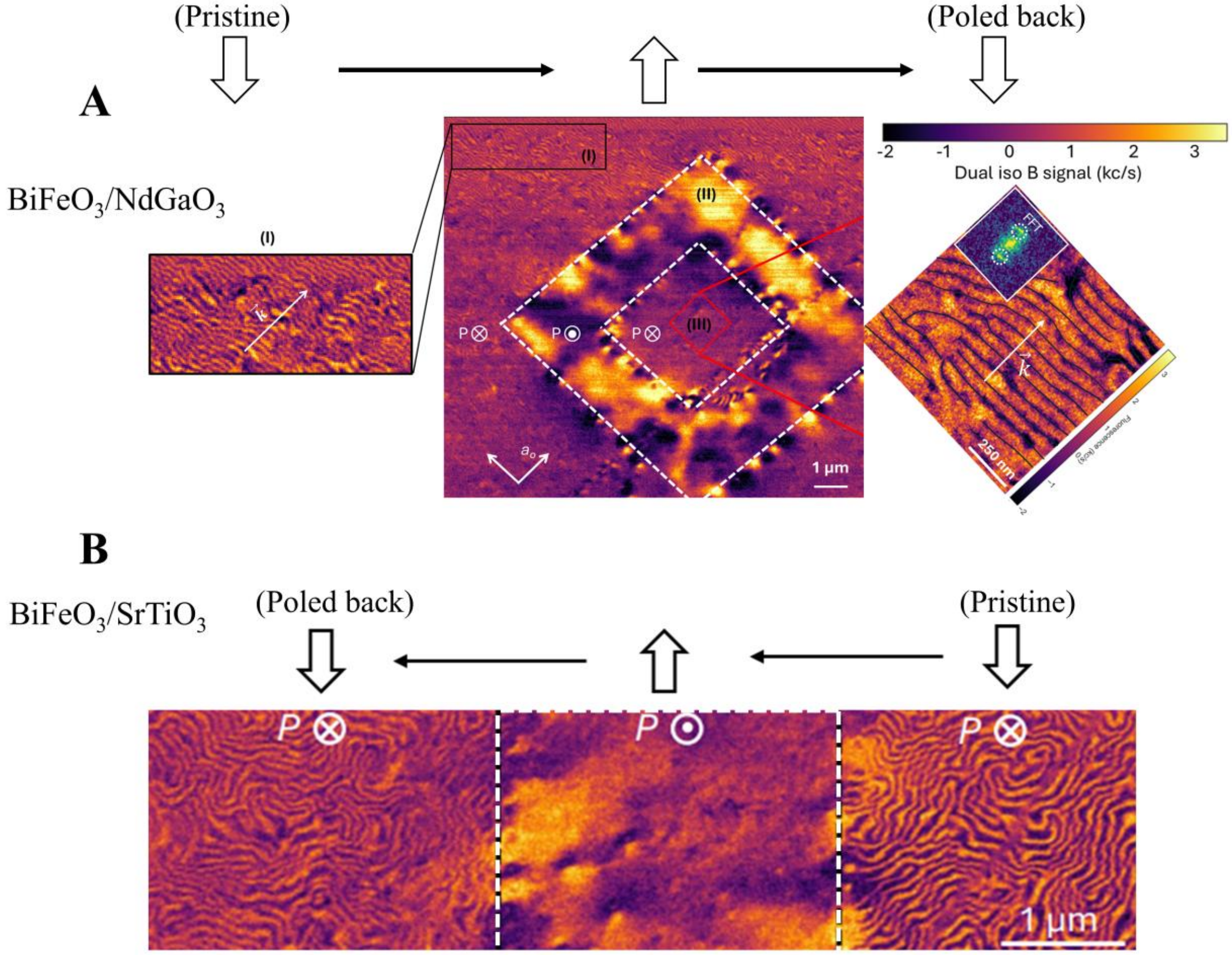

**Fig. S19. Large area scanning NV microscopy after box-in-a-box polarization switching via PFM tip.** The ferroelectric polarization of the as grown pristine down (↓) state is switched to up-polarization poled state (↑) (shown in the white dashed box) and subsequently poled back to down polarization state (↓) by an applied out-of-plane electric field via a PFM tip. **(A-B)**, NV scanning microscopy of the as-grown pristine region (showing robust single domain), switched region from down to up polarization state (showing homogenous signal without features under white dashed box), and the switched back region with down polarization (showing spin-cycloid on the surface of the film) of $(111)_{pc}$ $BiFeO_3$ on $NdGaO_3$ and $SrTiO_3$ respectively. Notably, after a complete switching cycle, the spin-cycloid comes back in the inside the poled back down-polarization region, which is important for practical device applications.

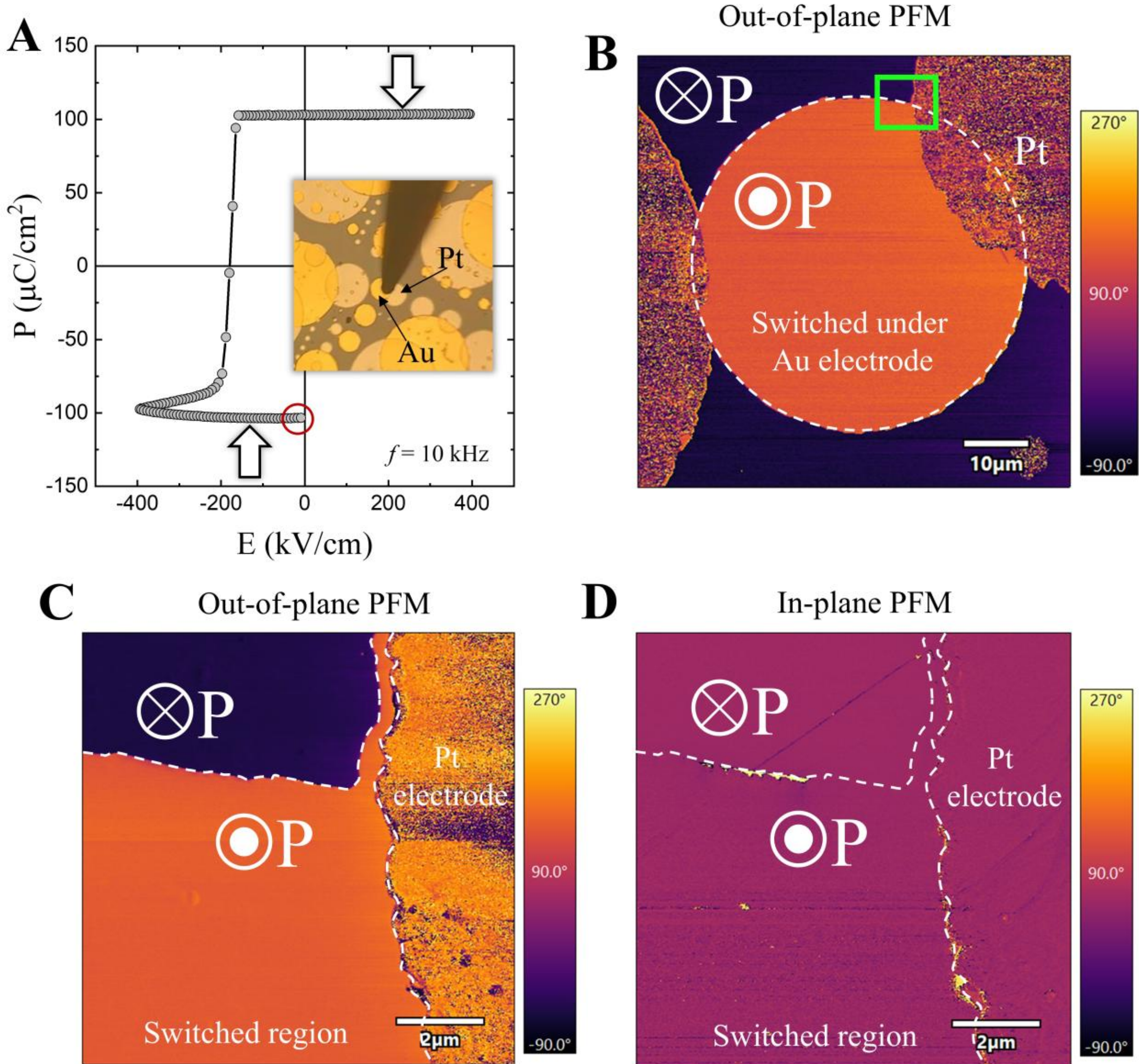

**Fig. S20. PFM signature of 180° polarization switching. (A)** Polarization vs electric field loop showing the polarization switching from down (↓) to up (↑) via a top Au electrode (inset). After switching the polarization to up state, the Au electrode is removed using KI solution. **(B)** Subsequent out-of-plane PFM image clearly showing the switched (up polarization) region under the electrode. **(C-D),** Out-of-plane and in-plane PFM scans of a boundary region (switched and unswitched regions) highlighted by a green box in **(B)**. There is clear 180º ferroelectric switching of the out-of-plane component without any intermediate 71° switched state.

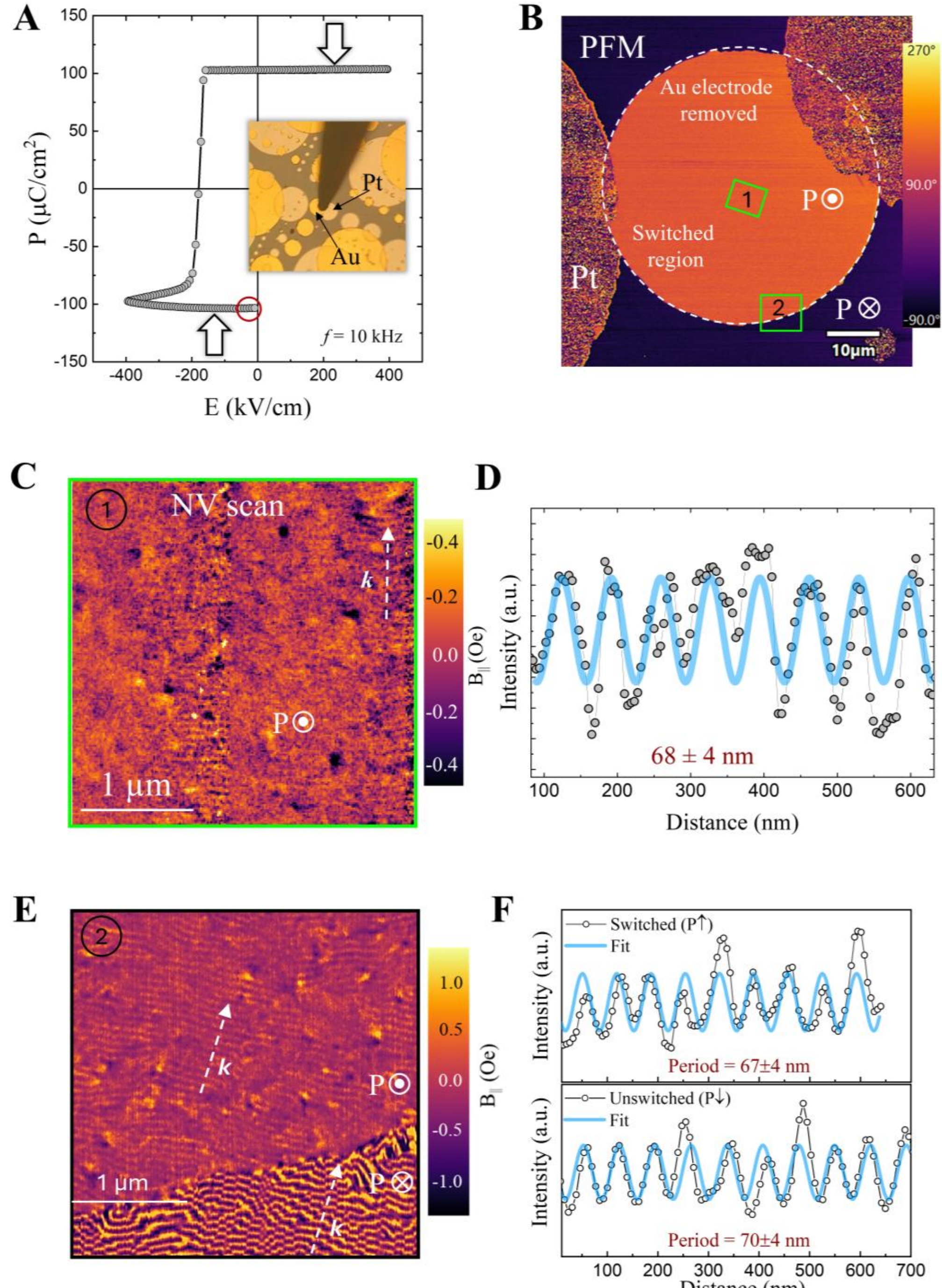

**Fig. S21. Scanning NV microscopy under a switched electrode with up polarization. (A)** Polarization vs electric field loop showing the polarization switching from down (↓) to up (↑) via a top Au electrode (inset). **(B)** Subsequent out-of-plane PFM image clearly showing the switched (up ↑ polarization) region under the electrode. **(C)** Corresponding scanning NV image on the switched (↑) region establishing clear presence of single spin-cycloid. **(D)** Line section profiles indicated by white dashed arrow highlighting similar spin-cycloid periodicity before and after polarization. (**E-F**) Similar NV scans on a boundary region highlighting robust spin cycloid in both the unswitched and switched regions. The data in **(E)** was acquired in QZabre quantum scanning system, Switzerland.

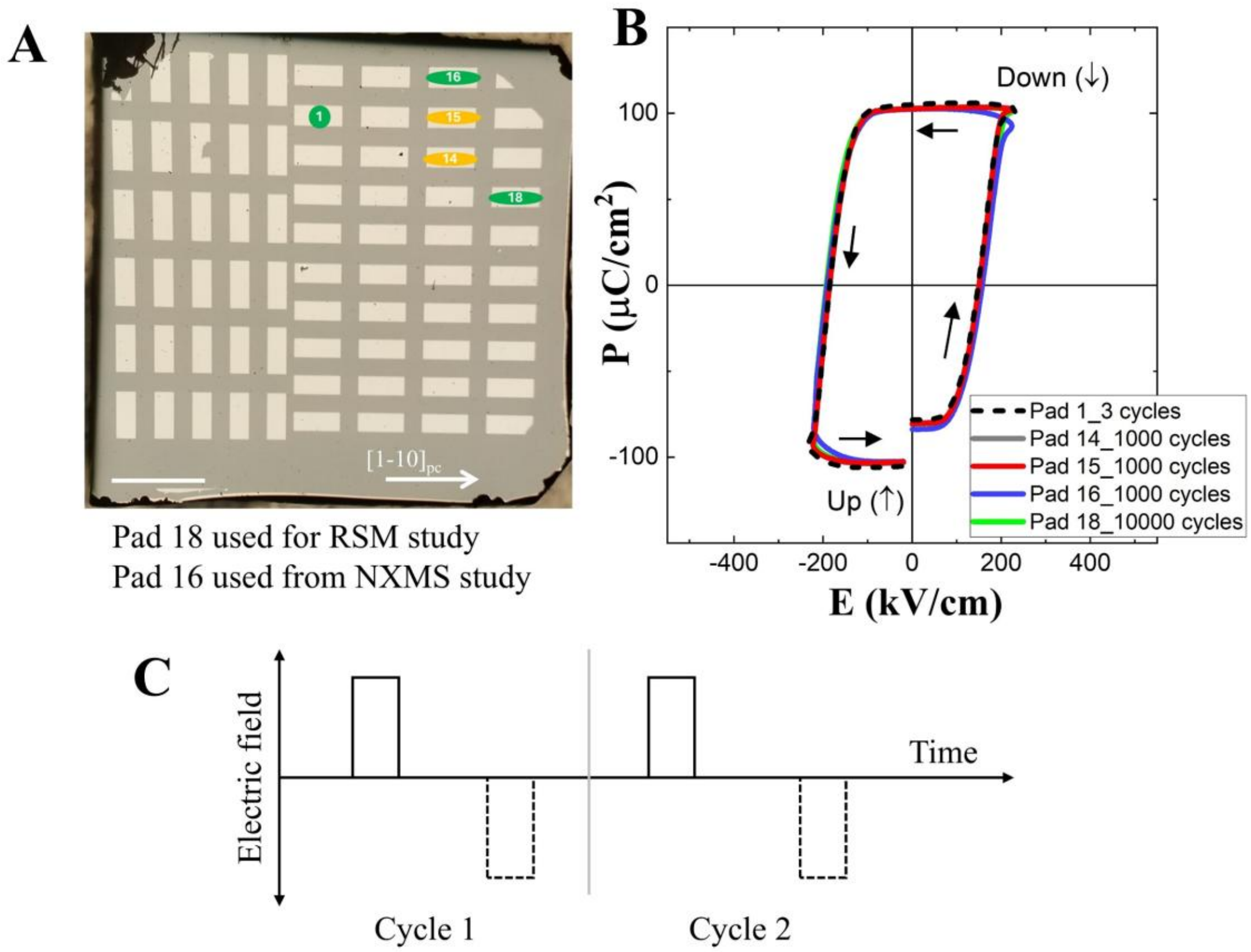

**Fig. S22. Sample and multiple cycles ferroelectric switching details. (A)** Optical microscopy image of the 1000 nm $BiFeO_3$ film grown on NdGaO3 $(011)_O$. **(B)** Ferroelectric Polarization vs Electric field loops after multiple cycles of electric field switching. Here, the final polarization state was in the up state. **(C)** Corresponding electric field pulse with a frequency of 200 Hz, used for switching the polarization by multiple times.

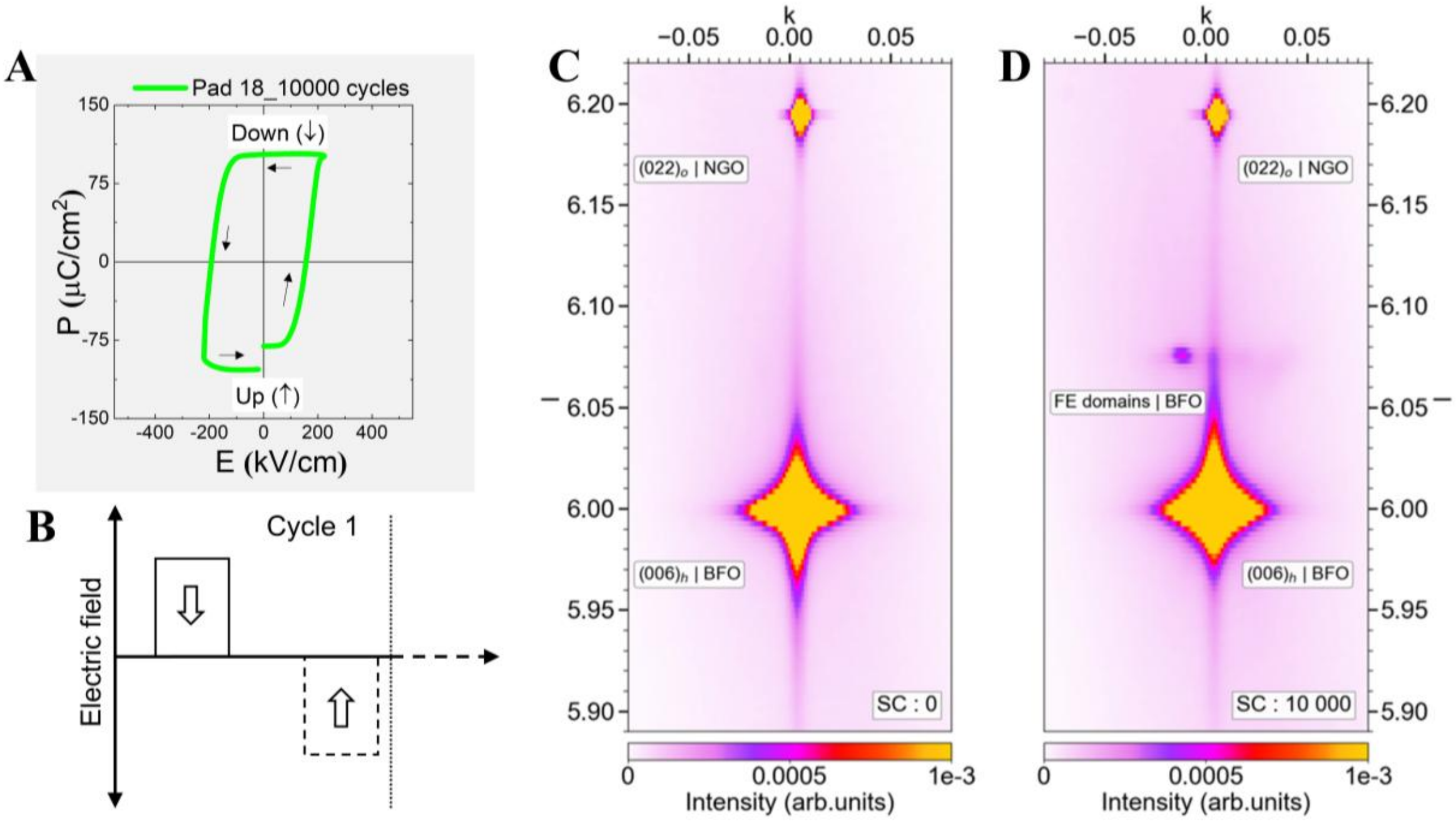

**Fig. S23. Robust single ferroelastic domain of $(111)_{pc}$ BFO grown on $NdGaO_3$ $(011)_O$. (A)** Ferroelectric Polarization vs Electric field loop after 10,000 cycles of electric field switching. Here, the final polarization state was in the up state. **(B)** Corresponding electric field pulse with a frequency of 200 Hz, used for switching the polarization multiple times. **(C)** RSM spectrum using synchrotron X-ray around the out-of-plane $(006)_h$ or $(111)_{pc}$ peak for BFO film grown on $NdGaO_3$ in the as-grown virgin state (where there is no electrode), which shows 100% single ferroelastic domain state. Refer to supplementary Fig.16 for hexagonal to cubic transformations. **(D)** Similar RSM spectrum under an electrode after 10,000 times electric field switching (under pad 16, shown in Fig. S22), shows a very nominal degradation of a single ferroelastic domain by only ~0.4%, which is only detectable under a synchrotron x-ray beam.

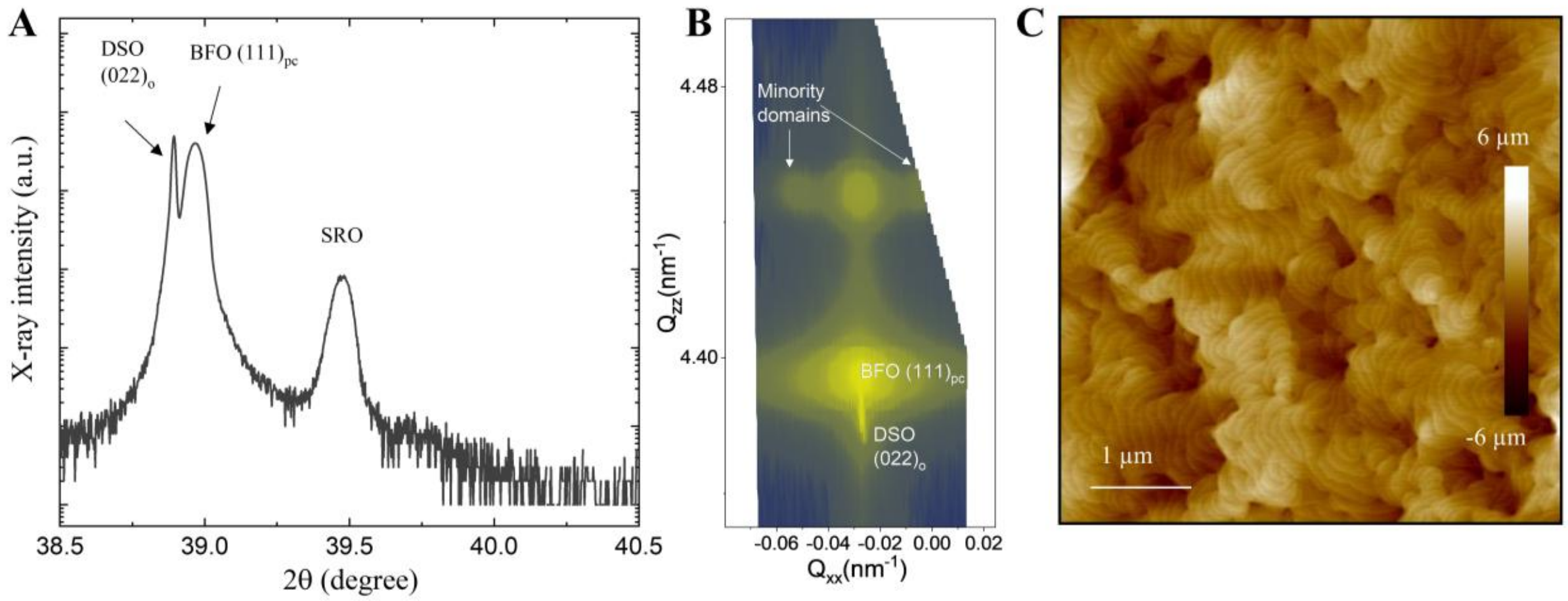

**Fig. S24. Structural characterization of $(111)_{pc}$ $BiFeO_3$ film grown on $DyScO_3$ $(011)_O$. (A)** Room temperature XRD spectrum of 1000 nm $(111)_{pc}$ $BiFeO_3$ (BFO) thin films grown on orthorhombic $(011)_O$ $DyScO_3$ (DSO) substrate. **(B)** Corresponding reciprocal space mapping indicates seeding of undesired minority domains in the as-grown state. **(C)** Atomic force microscopy image displays an atomically smooth surface with a root-mean-square (RMS) roughness of ~4Å.

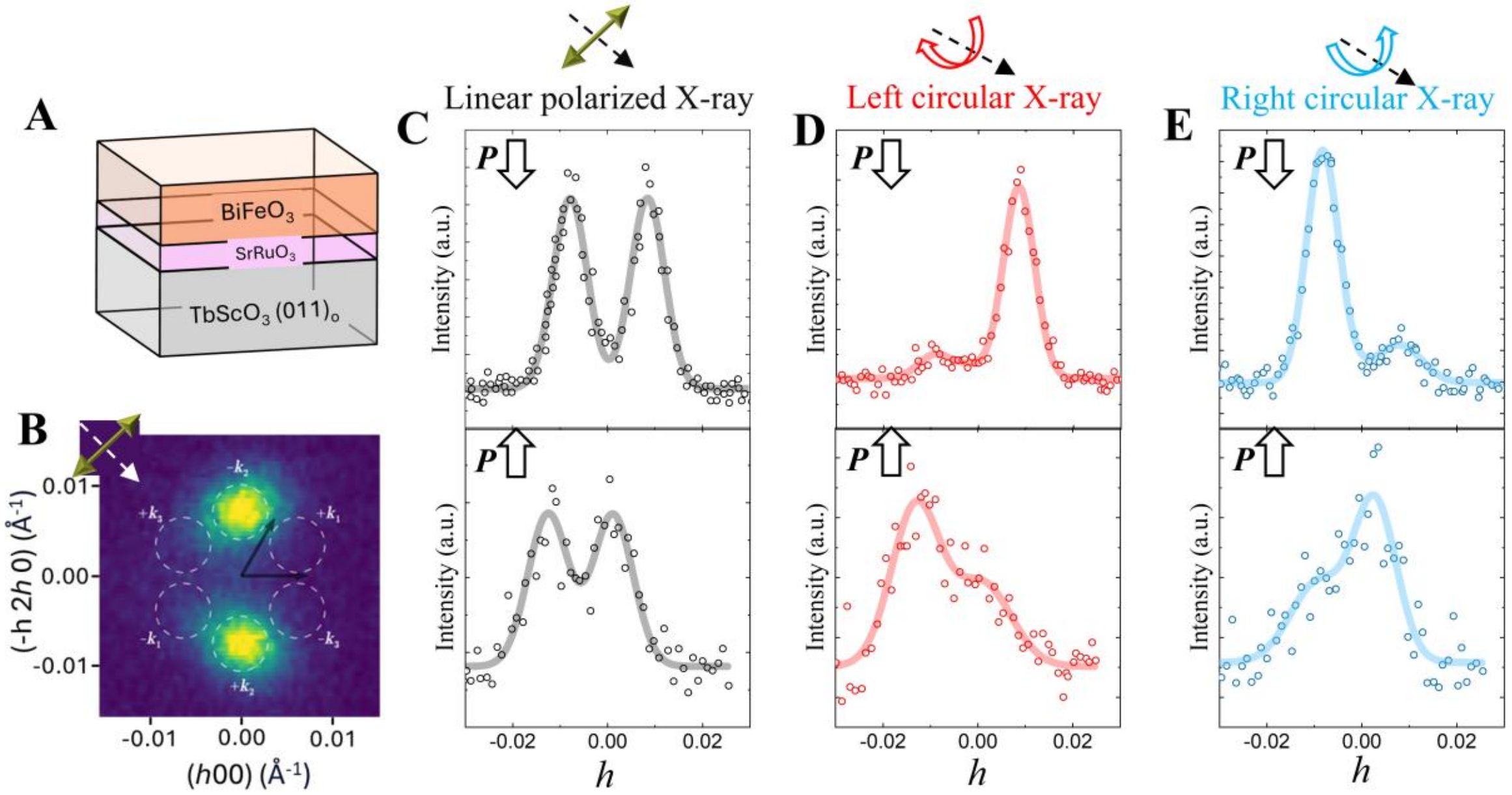

**Fig. S25. Mapping of spin-cycloid before and after switching of ferroelectric polarization under different polarizations of X-ray beam. (A)** Schematic structure of $(111)_{pc}$ $BiFeO_3$ grown on $TbScO_3$ $(011)_O$ substrate buffered by a $SrRuO_3$ layer. **(B)** A typical NXMS scan using linearly polarized X-ray of the AFM spin cycloid showing single variant. (**C-E**) Line section profiles of the NXMS scans before and after electrical switching of ferroelectric polarization under linearly polarized, left circularly polarized, and right circularly polarized incident x-rays, respectively. Here, it is very clear that upon switching the cycloid changes its polarity as observed from circularly polarized x-ray scans. These data were reproduced from our previous work(*4*).

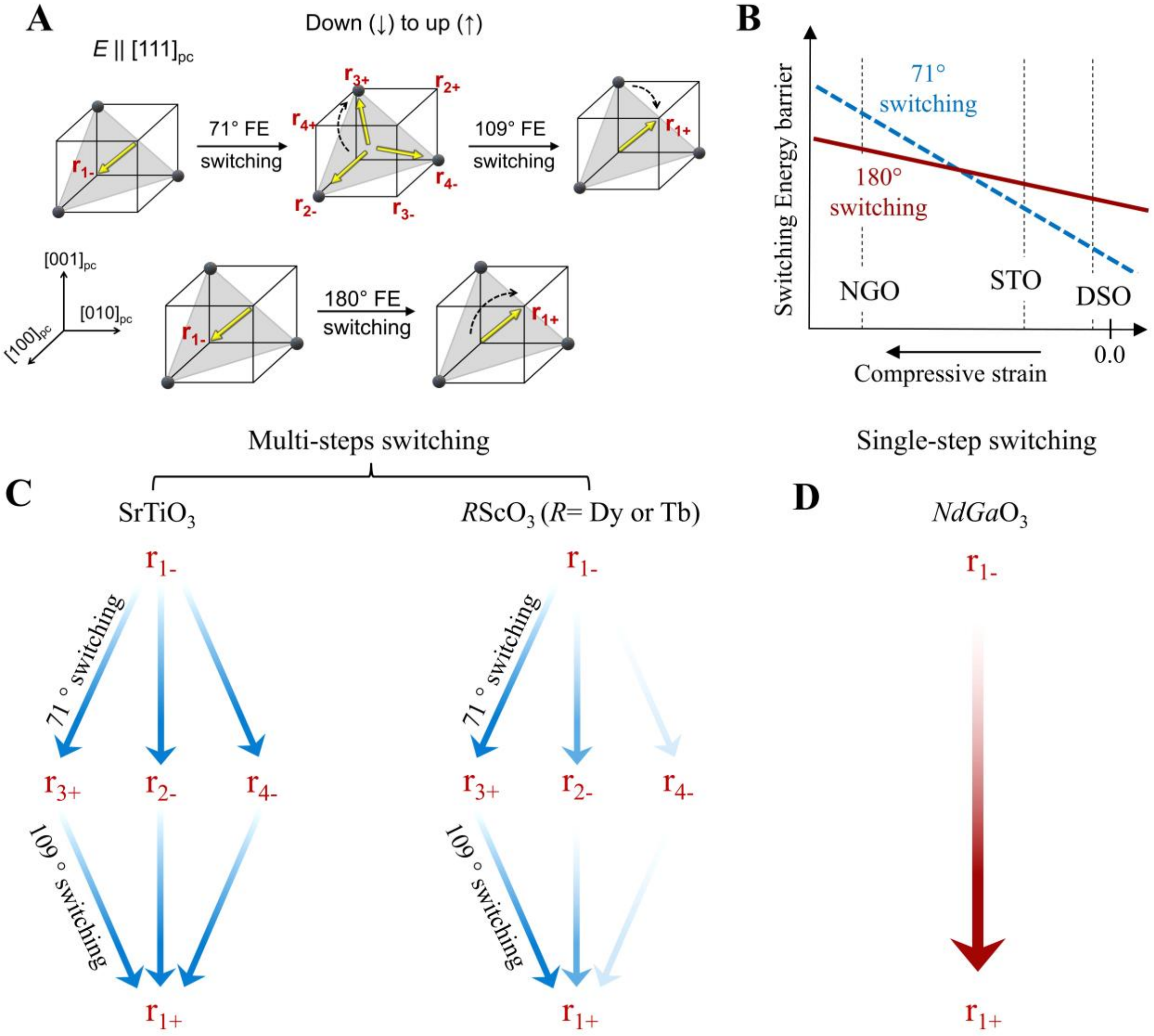

**Fig. S26. Mechanism of deterministic ferroelectric switching (A)** Schematic representation of multi-stage 71° and single step 180° ferroelectric polarization switching in $[111]_{pc}$ BFO. **(B)** Qualitative energy stability between multi-stage 71° and single-step 180° ferroelectric polarization switching in $[111]_{pc}$ BFO(*19*), with increasing compressive strain. It is expected that while 71° switching is facilitated in $[111]_{pc}$ BFO grown on $SrTiO_3$ (STO) and $DyScO_3$ (DSO), 180° deterministic switching can be achieved when grown with high-compressive strain (e.g., $NdGaO_3$ (NGO)). **(C)** The presence of equally favorable multi-path 71° switching in isotropically strained BFO on STO and *anisotropic-tensile* strained DSO, leading to the formation of minority domains (along the tensile strain direction) and charged domain walls leading to polarization fatigue. **(D)** Single step 180° ferroelectric polarization switching in $[111]_{pc}$ BFO on NGO $(011)_O$ as obtained from phase-field simulations that can lead the fatigue-free ferroelectric switching response.

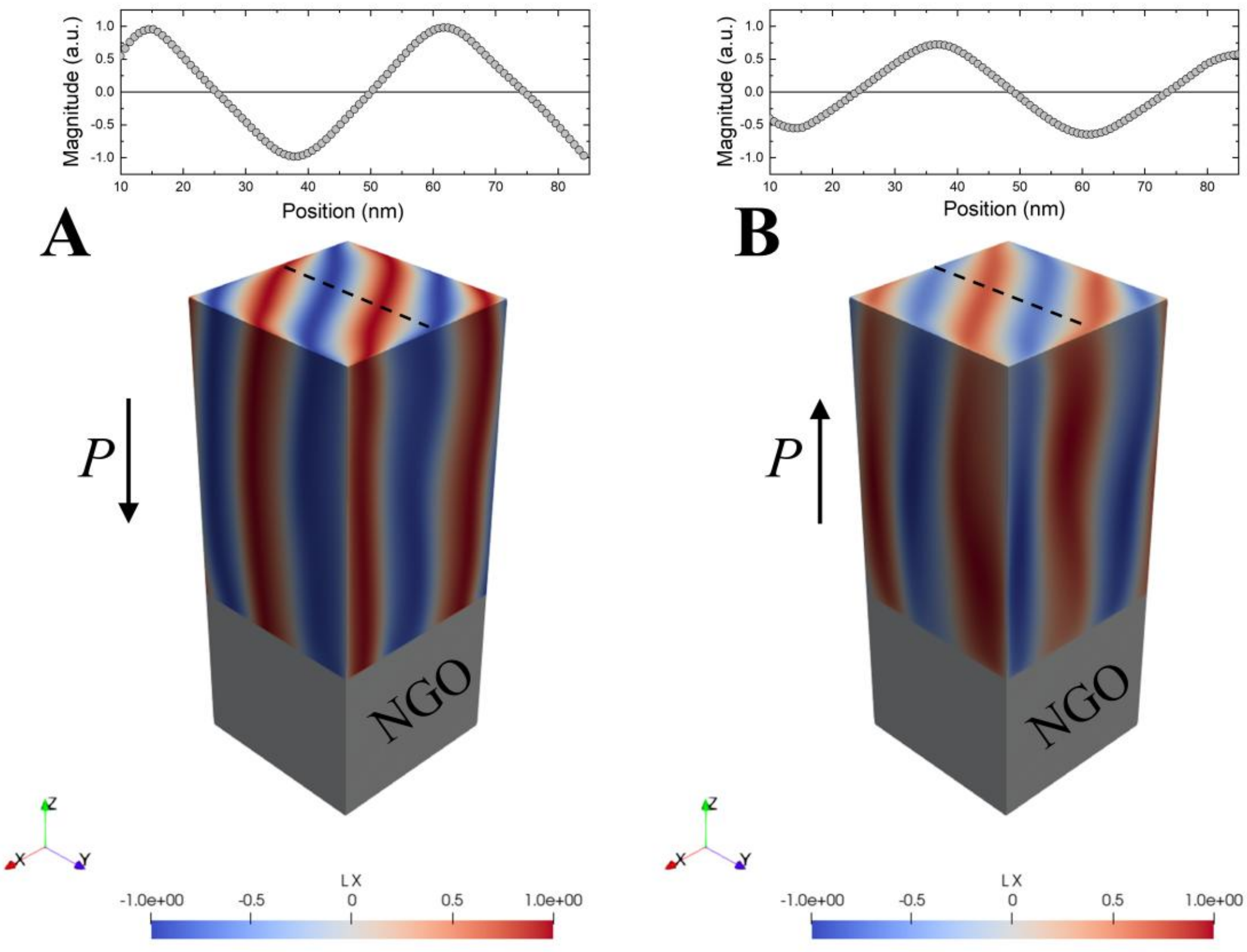

**Fig. S27. (A-B)** Phase field simulation of the single domain structure before and after ferroelectric polarization switching.

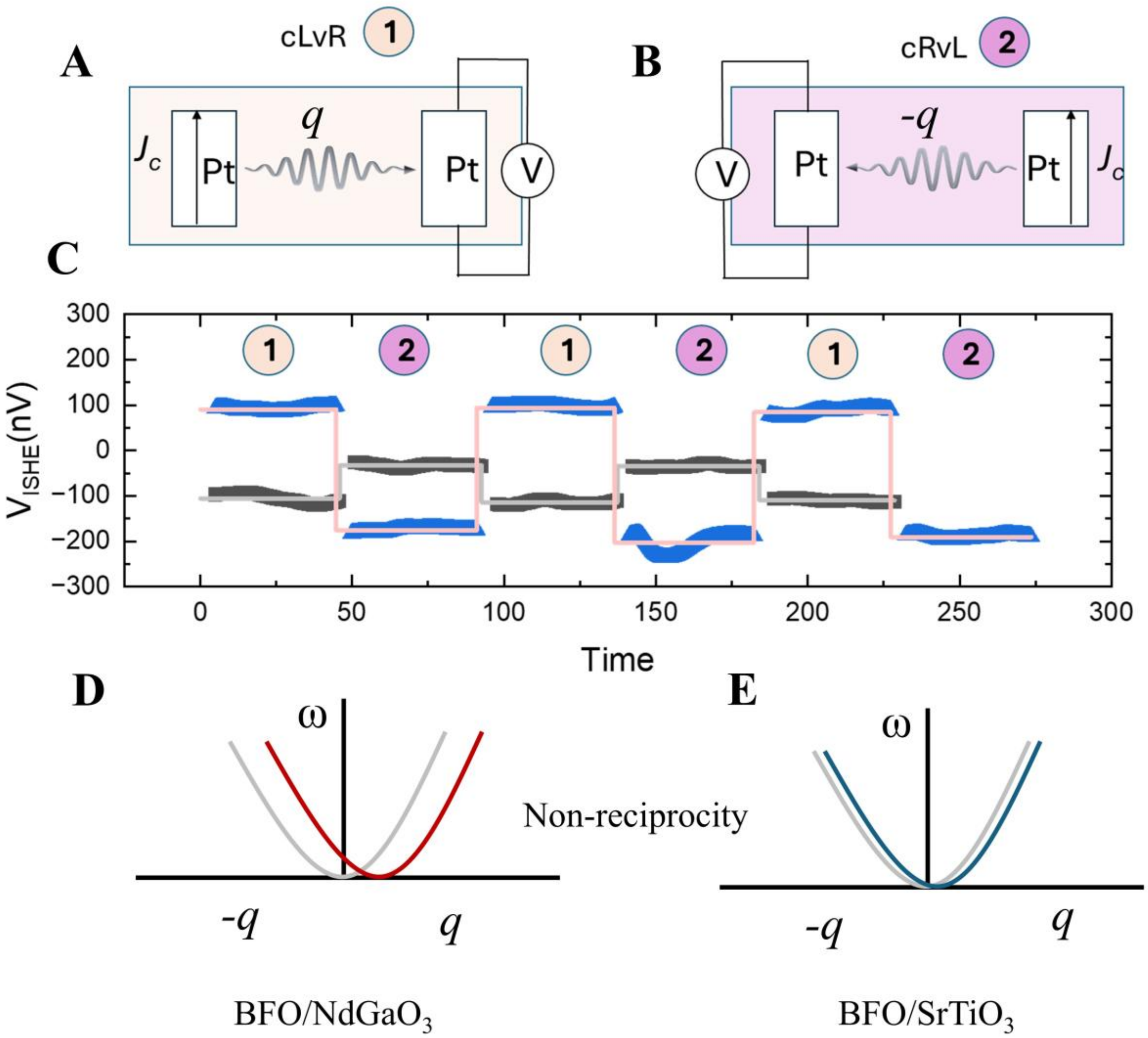

**Fig. S28. Non-local spin transport in single vs multiple spin cycloids AFM domain in $(111)_{pc}$ BFO. (A)** current left voltage right (cLvR) and **(B)** current right voltage left (cRvL). **(C)** Non-local data measured in the (111) BFO deposited on $SrTiO_3$ (black data) and on $NdGaO_3$ (blue data). Line is the average fit to the data. **(D-E)** Non-reciprocity in when the thermal magnons flow in opposite direction. The non-reciprocity is reduced in case of $SrTiO_3$ due to homogeneous nature of the spin cycloid.

Non-reciprocity refers to an asymmetry of responses or signals between forward and backward directions. Previously, it has been observed that such non-reciprocity can occur during magnon transport in case of $(001)_{pc}$ BFO thin film(*30*), that has been qualitatively interpreted due to magnon band splitting caused by Dzyaloshinskii–Moriya interactions.